\PassOptionsToPackage{hyphens}{url}
\documentclass{sig-alternate}
\usepackage{mathptmx}

\usepackage{authblk}
\usepackage{fancyhdr}
\usepackage[normalem]{ulem}
\usepackage[hyphens]{url}
\usepackage[sort,nocompress]{cite}
\usepackage[final]{microtype}
\usepackage[dvipsnames]{xcolor}
\usepackage[bookmarks=true,breaklinks=true,letterpaper=true,colorlinks,linkcolor=black,citecolor=blue,urlcolor=black]{hyperref}

\usepackage{booktabs}
\usepackage{xspace}
\usepackage{enumitem}
\usepackage[binary-units=true]{siunitx}
\usepackage{amsmath}
\usepackage{ifthen}
\usepackage{graphicx}
\usepackage{anyfontsize}
\usepackage{multirow}
\usepackage{listings}
\usepackage{setspace}
\usepackage{float}
\usepackage[us,12hr]{datetime}
\usepackage{tikz}
\usepackage{dblfloatfix}
\usepackage[keeplastbox]{flushend}

\fancypagestyle{firstpage}{
  \fancyhf{}
  
  \fancyhead[C]{} 
  \fancyfoot[C]{\thepage}
}

\newcommand{\paperlinespacing}[1]{0.7}

\DeclareSIUnit{\flop}{FLOP}

\newcommand{\titleShort}{Mensa\xspace}
\newcommand{\accelA}{Pascal\xspace}
\newcommand{\accelB}{Pavlov\xspace}
\newcommand{\accelC}{Jacquard\xspace}

\newcommand{\paratitle}[1]{\vspace{4pt}\noindent\textbf{#1.}}
\newcommand{\paratitleattop}[0]{\vspace{-4pt}}

\newcommand{\todo}[1]{}
\newcommand{\todob}[1]{}
\newcommand{\am}[1]{#1}
\newcommand{\sg}[1]{{#1}}
\newcommand{\sgii}[1]{{#1}}

\newcommand{\affilCMU}{$^\dag$}
\newcommand{\affilGoogle}{$^\odot$}
\newcommand{\affilETH}{$^\diamond$}
\newcommand{\affilUIUC}{$^\ddag$}

\title{\vspace{-16pt}Mitigating Edge Machine Learning Inference Bottlenecks: \\ An Empirical Study on Accelerating Google Edge Models}
\author{\vspace{-48pt}%
\large{Amirali Boroumand\affilCMU\qquad%
Saugata Ghose\affilUIUC\qquad%
Berkin Akin\affilGoogle\qquad%
Ravi Narayanaswami\affilGoogle}%
\\\vspace{-8pt}%
{\large Geraldo F. Oliveira\affilETH\qquad%
Xiaoyu Ma\affilGoogle\qquad%
Eric Shiu\affilGoogle\qquad%
Onur Mutlu\affilETH\affilCMU}%
\\\vspace{5pt}%
{\fontsize{11}{12}\selectfont\it%
\affilCMU Carnegie Mellon University\qquad%
\affilUIUC University of Illinois at Urbana--Champaign\qquad%
\affilGoogle Google\qquad%
\affilETH ETH Z{\"u}rich%
\vspace{-8pt}}}

\begin{document}
\sloppy
\maketitle
\thispagestyle{firstpage}
\pagestyle{plain}

\setstretch{0.99}

\begin{abstract}

\sg{As the need for edge computing grows, many modern consumer devices now
contain edge machine learning (ML) accelerators that can compute a wide range of neural network (NN) models while still fitting within tight resource constraints.}
We analyze 
\am{a commercial Edge TPU} using 24 Google edge NN models \sg{(including CNNs, LSTMs, transducers, and RCNNs)}, and find that 
\sg{the accelerator} suffers
from three shortcomings,
\sg{in terms of computational throughput, energy efficiency, and memory access handling.}
We comprehensively study the characteristics of each \sg{NN} layer in all of the Google edge models, and find that these shortcomings arise from the one-size-fits-all approach of \sg{the accelerator}, as there is a high amount of heterogeneity in key layer characteristics both across different models and across different layers in the same model.

\sg{We} propose a new \sg{acceleration} framework called Mensa. Mensa incorporates multiple heterogeneous ML \sg{edge} accelerators \am{(including both on-chip and near-data accelerators)}, each of which caters to the characteristics of a particular subset of models. At runtime, Mensa schedules each layer to run on the best-suited accelerator, accounting for both efficiency and inter-layer dependencies. 
\sg{As we analyze the Google edge NN models,} we discover that all of the layers naturally group into a small number of clusters, 
\sg{which allows us to design an efficient implementation of Mensa for these models with only three specialized accelerators.}
\am{Averaged across all 24 Google edge models, Mensa improves energy efficiency and throughput by 3.0x and 3.1x
over the Edge TPU, and by 2.4x and 4.3x over Eyeriss~v2, a state-of-the-art accelerator}.

\end{abstract}


\section{Introduction}  
\label{sec:intro}

Modern consumer devices make widespread use of
machine learning (ML).
The growing complexity of these devices, combined with 
increasing demand for privacy, connectivity, and real-time responses,
has spurred significant interest in pushing ML inference computation to
the edge (i.e., into these devices, instead of in the cloud)~\cite{edge-facebook, edge-nature,eyerissv2}.
Due to the resource-constrained nature of edge platforms, 
they now employ specialized
energy-efficient accelerators for on-device inference
(e.g., Google Edge TPU~\cite{edge-tpu}, NVIDIA Jetson~\cite{jetson}, 
Intel Movidius~\cite{movidius}).
At the same time, neural network (NN) algorithms are evolving rapidly,
which has led to \sg{many types of}
NN models
(e.g., CNNs~\cite{simonyan2015very}, LSTMs~\cite{lstm-google},
GRUs~\cite{gru}, Transducers~\cite{transducer, transducer2},
hybrid models~\cite{lrcn,rcnn-google}), 
each targeting various applications 
(e.g., face detection~\cite{simonyan2015very},
speech recognition~\cite{transducer,transducer3, lstm-google}, 
translation~\cite{google-translation}, 
image captioning~\cite{rcnn-google,lrcn}).

\sg{Google's state-of-the-art Edge TPU~\cite{edge-tpu} provides an optimized one-size-fits-all solution
across this wide variety of NN models, while being mindful of edge device area
and energy constraints.}
Unfortunately, 
\sg{we find that it is very challenging to simultaneously achieve}
high energy efficiency (\si{\tera\flop\per\joule}),
computational throughput (\si{\tera\flop\per\second}), and
area efficiency (\si{\tera\flop\per\milli\meter\squared})
for each workload \sg{with this one-size-fits-all approach}.
We conduct an in-depth analysis of inference execution
 on \am{a commercial Edge TPU},
across 24 state-of-the-art Google edge models spanning four popular NN model types:
(1)~CNNs, 
(2)~LSTMs~\cite{lstm-google}, 
(3)~Transducers~\cite{transducer, transducer3, transducer4}, and 
(4)~RCNNs~\cite{rcnn-google,lrcn}.
\sg{These models are used in several Google mobile applications,
such as image classification, object detection, semantic segmentation,
automatic speech recognition, and image captioning.}
Based on our analysis (Section~3), we find that the accelerator suffers from
 three major shortcomings. First, 
\sg{the accelerator utilizes only 1/4 of its peak throughput,
averaged across all models (and less than 1\%
utilization for LSTMs and Transducers, the worst case)}.
Second, despite using specialized logic, the accelerator
\sg{provides only 37\% of} its theoretical peak energy efficiency (\si{\tera\flop\per\joule}) on average.
Third, the accelerator's memory system is often \sg{a large bottleneck}.
As an example, while large on-chip storage buffers
(e.g., several megabytes) account for a significant portion of energy consumption
(e.g., 48.1\% static and 36.5\% dynamic energy during CNN inference),
they are often ineffective \sg{at} reducing off-chip accesses, and cannot accommodate
the parameters of larger NN models.

To identify the root cause of these shortcomings, we perform \sg{the first} comprehensive 
per-layer analysis of \sg{the Google edge NN models}, and make two key observations.
First, there is significant variation in terms of
 layer type, shape, and characteristics (e.g., \si{\flop\per{Byte}} ratio, footprint, intra- and inter-layer
 dependencies)
\emph{across} the models. 
For example, Transducer layers differ drastically (by as much as two orders 
of magnitude) from CNN layers in terms of parameter footprint and \si{\flop\per\byte}.
Second, even \emph{within each model}, there is high variation in terms of 
layer types and shapes 
(e.g., pointwise, depthwise, fully-connected, standard convolution, recurrent).
This leads to up to two orders of magnitude of variation for
layer characteristics within a single model.
We \sgii{quantify for the first time how} intra-model \sg{variation} is dramatically higher in
edge models \sg{compared to previously-studied}
traditional models (e.g., \cite{alex-net,simonyan2015very}), as edge models employ
several techniques (e.g., separable convolutions~\cite{squeezenet,mobilenet})
to reduce computational complexity and layer footprint,
in order \sg{to optimize the models for} resource-constrained edge devices.

\sg{Despite this large variation, many}
state-of-the-art edge ML
 accelerators~\cite{edge-tpu, movidius, movidius-arch, jetson, tangram,brainwave} take a monolithic design approach, where they
 equip the accelerator with a large PE array, large on-chip buffers, and a fixed
 dataflow (e.g., output stationary). While this approach might work for
 a specific group of layers (e.g., traditional convolutional layers with high compute intensity
 and high data reuse), we find that it 
\sg{leads to throughput and energy efficiency shortcomings}
across the \sg{significantly more} diverse edge NN models,
\sg{as illustrated by two examples}.
\sg{First, in the memory system,} state-of-the-art accelerators employ highly overprovisioned 
on-chip buffers
that are unable to
 effectively reduce off-chip parameter traffic and provide
 high bandwidth to PEs, leading to PE underutilization. 
\sg{Second, state-of-the-art accelerators use a fixed dataflow across all layers.}
Due to the drastic variation across different layers,
the fixed dataflow often misses spatial/temporal reuse opportunities across layers.

\sg{A number of recent works~\cite{eyerissv2,maeri} cater to NN variation by enabling reconfigurability for
parts of the accelerator.  For example, Eyeriss~v2~\cite{eyerissv2} provides 
the ability to reconfigure the on-chip interconnect and make use of a smaller
PE array.  Unfortunately, as models become more diverse and go beyond the
structure of more traditional CNNs, reconfigurable accelerators face 
three issues:
(1)~they do not provide the ability to reconfigure a number of essential
design parameters (e.g., on-chip buffers, memory bandwidth);
(2)~they can require frequent \sg{online} reconfiguration to cater to increasing intra-model
heterogeneity, \sg{with associated overheads}; and
(3)~they \sg{make it difficult to co-design} the dataflow with key components such as the memory system.}
\am{The \emph{key takeaway} from our \sg{extensive analysis of Google edge NN models on the Edge TPU}
is that \emph{all key components} 
of an edge accelerator (i.e., PE array, dataflow, memory system)
must be co-designed based on specific layer characteristics to achieve high utilization and energy efficiency.}
Our goal is to revisit the design of edge ML accelerators such that they are aware of and can \sg{fully} exploit
the \sg{growing} variation within and across edge NN models.

\am{To this end, 
we} propose \titleShort, the first general HW/SW composable
framework for ML acceleration in edge devices.
\sg{The key idea of \titleShort is to incorporate and manage layer
execution across multiple on-chip and near-data accelerators,
each of which is small and tailored to the characteristics of a particular
subset of layers.
As we study the characteristics of different layers in our edge models,
we make an important discovery: the layers naturally group into a small number of
clusters that are based on a subset of these characteristics.
This allows us to limit the number of different accelerators in a
\titleShort design to the number of clusters
(five for our Google edge models).
We design a runtime for \titleShort to determine which of these
accelerators should execute each of the layers,
using information about
(1)~which accelerator is
best suited to the layer's characteristics, and 
(2)~inter-layer dependencies.}

\sg{Using our insight about layer clustering, we develop one possible design for \titleShort that is
optimized for our Google edge workloads.
We find that our accelerator designs should center around two
key layer characteristics (memory boundedness, and 
activation/parameter reuse opportunities).
Of our five clusters, Clusters 1 and 2 have similar key characteristics,
as do Clusters 4 and 5, allowing us to reduce the number of accelerators
needed in our example implementation to three (which we call
\accelA, \accelB, and \accelC).}
\sg{\accelA (for compute-centric layers)}
uses a dataflow that enables temporal reduction of output activations 
and spatial multicasting of parameters.  This dataflow allows us to
design a memory system with an on-chip buffer that is 16x smaller than
\am{our baseline Edge TPU},
and greatly reduces on-chip network
traffic, while still keeping the processing elements (PEs) highly utilized.
\sg{\accelB (for LSTM-like data-centric layers)}
uses a dataflow that enables temporal reduction of output activations,
and identifies opportunities to schedule the parallel execution of layer operations
in a way that improves parameter reuse,
reducing off-chip
parameter traffic.
\sg{\accelC (for other data-centric layers)}
uses a dataflow that exposes reuse opportunities in parameters and
reduces the size of the parameter buffer.
\sg{We provide both of our data-centric accelerators with high
memory bandwidth by placing them in the logic layer of 3D-stacked memory,}
and use significantly smaller PE arrays for the accelerators compared to
the PE array in \accelA.

Our evaluation shows that compared to \am{our baseline Edge TPU},
    \titleShort reduces total inference energy by 66.0\%,
 improves energy efficiency (\si{\tera\flop\per\joule}) by 3.0x, 
    and increases computational throughput (\si{\tera\flop\per\second}) by 3.1x
(averaged across all 24 Google edge NN models). \am{\titleShort improves inference energy 
efficiency and throughput 
by 2.4x and 4.3x over Eyeriss~v2, a state-of-the-art accelerator}.

We make the following \textbf{contributions} in this work:

\begin{itemize}[leftmargin=1em,nosep]

\item We conduct the first in-depth analysis of how \sg{the Google Edge TPU operates}
across a wide range of state-of-the-art \sg{Google} edge NN models.
Our analysis reveals three key shortcomings of \am{the Edge TPU}
\sg{for these models}:
(1)~\sg{poor throughput due to} significant PE underutilization,
\sg{(2)~\am{low energy efficiency}, and
(3)~a large memory bottleneck.}

\item We comprehensively analyze the key characteristics of each layer in 
\sg{24 Google} edge NN models. We make \sgii{three} observations from our analysis:
(1)~layer characteristics vary significantly both \emph{across} models
and across layers \emph{within a single model},
(2)~the monolithic design of \am{state-of-the-art accelerators (e.g., the Edge TPU)} is the
root cause of their shortcomings, and
\sgii{(3)~layers naturally group into a small number of clusters based on their
characteristics}.

\item We propose \titleShort, a new framework for efficient edge ML acceleration.
\sg{\titleShort is the first accelerator to exploit the significant compute and memory
demand heterogeneity that we observe in state-of-the-art edge NN models,
through the use of a runtime scheduler and a few small, carefully specialized
accelerators.}

\item
We \sg{create a \titleShort design for our Google edge models,
and find that it} is significantly more energy efficient and \sg{provides higher throughput}
than \am{a commerical Edge TPU} \sg{and Eyeriss v2, a state-of-the-art ML accelerator.} 

\end{itemize}


\section{Background}
\label{sec:bkgd}

\paratitleattop{}
\paratitle{CNNs} 
Convolutional neural networks (CNNs) are 
feed-forward models that successfully 
capture spatial features.
A CNN is mostly composed of convolutional layers. Each convolutional layer
(1)~performs a 2D convolution operation between the input activations and parameters from kernels, 
and (2)~passes the result
through a non-linear activation function (e.g., ReLU, sigmoid, tanh) to produce the output activations.  

\paratitle{LSTMs} 
Long short-term memories (LSTM) are a class of NNs 
 that are effective for sequence-to-sequence application domains.
 An LSTM
contains recurrent connections that allow the model to propagate and maintain context information across
the input sequence. At each step, an LSTM makes
 a prediction based on the current input ($x_{t}$) and activations from the previous time step
 ($h_{t-1}$). 

The LSTM network consists of
 multiple LSTM layers, where each includes several LSTM cells. 
 Each LSTM cell has \emph{four gates} (input, forget, update, and output) that allow the
 cell to regulate the information flow and update the cell state accordingly. 
 At each step, the cell receives the input ($x_{t}$) and input hidden
 vector ($h_{t-1}$), which is the output hidden vector from the previous
 cell (i.e., the recurrent connection). Each gate performs two
 matrix-vector multiplies (MVM): (1)~an input MVM with the input parameter 
matrix ($W_{x}$) and the input vector ($x_{t}$), and (2)~a hidden MVM with
the hidden parameter matrix ($W_{h}$) and the input hidden vector ($h_{t-1}$).

\paratitle{Transducers}
Transducers are the state-of-the-art NN model for end-to-end automatic speech recognition
 systems~\cite{transducer,transducer2, transducer3,transducer4}.
A transducer has three major blocks: (1)~an encoder, which receives acoustic features and converts
 them into a high-level representation; (2)~a prediction network, which generates linguistic outputs (i.e., high-level representation) 
that depend on the entire sequence of labels; 
 and (3)~a joint, which is a feed-forward joint that receives inputs from
both the encoder and a prediction network that depends only on
label histories.  
The encoder, prediction, and joint blocks are typically implemented by stacking several LSTM layers. 

\paratitle{RCNNs}
Recurrent convolutional neural networks (RCNNs; also known as LRCN~\cite{lrcn}) are a class of hybrid NN models that attempt to capture 
spatio-temporal information~\cite{rcnn-google, rcnn-caption, dnpu,lrcn-fpga,crnn}. RCNNs leverages the success of CNNs in extracting
 spatial features, and combine them with LSTM-based models that excel at
 detecting temporal relationships.
RCNNs typically employ multiple convolutional layers in the front end of the network to perform
 feature extraction on input data, and then pass the CNN layer output to an
 LSTM-based model that performs sequence prediction.


\section{TPU \& Model Characterization}
\label{sec:motiv}

\label{sec:motiv:methodology}

We analyze 24 Google edge models (including CNNs, LSTMs, Transducers, and RCNNs)
using \am{a commercial Edge TPU as our baseline accelerator}.
\am{The Edge TPU has a generic tiled architecture, similar to other state-of-the-art accelerators~\cite{eyeriss, tetris, tangram, tpu}.
It includes a 2D array of PEs (64x64), 
 where each PE has a small register file 
 to hold intermediate results. 
 The accelerator has two large SRAM-based on-chip buffers to
 hold model parameters and activations~\cite{edge-tpu-compiler}.}

\subsection{Google Edge TPU Shortcomings}
\label{sec:motiv:accelerator}

\begin{figure}[t]
    \centering
        \centering
        \includegraphics[width=\linewidth]{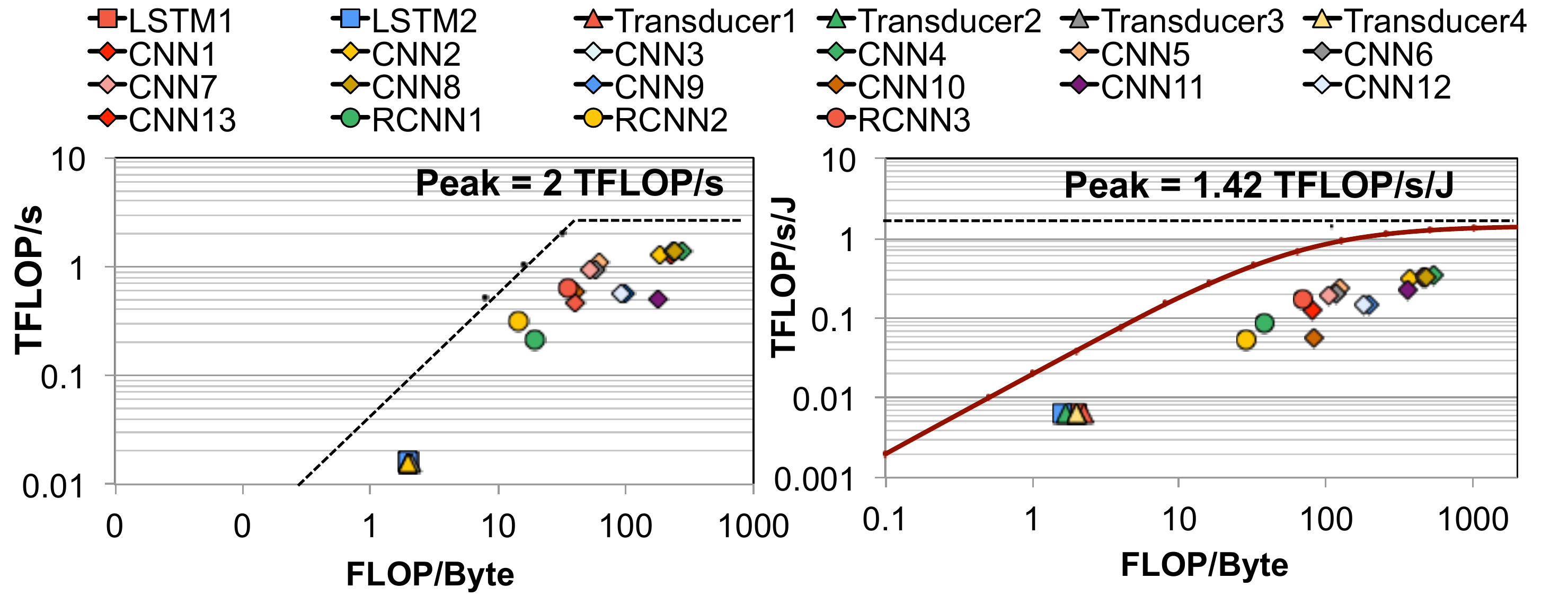}%
\vspace{-24pt}
    \caption{Throughput roofline (left) and energy roofline (right) for \am{the Edge TPU} across all models.
        }
    \label{fig:roofline-throughput}
    \label{fig:roofline-energy}
\end{figure}

\textbf{1. The accelerator often suffers from extreme underutilization of the PEs.}
The \am{Edge TPU} has a theoretical peak throughput of \SI{2}{\tera\flop\per\second},
 which is orders of magnitude higher than CPUs and GPUs. However, 
the accelerator operates \emph{much} lower than
 peak throughput during inference execution (75.6\% lower on average).
Figure~\ref{fig:roofline-throughput} (left) shows the
 roofline model of throughput for \am{the Edge TPU},
along with the \sg{measured} throughput of all of our edge models. 
The PE utilization is consistently low across all models.
Transducer and LSTM models have the most underutilization,
with both achieving less than 1\% of peak throughput.
While CNN and RCNN models do somewhat better,
they achieve only 40.7\% of peak utilization on average
(going as low as 10.2\%).

\textbf{2. Despite using specialized logic, \am{the Edge TPU} operates far below its
theoretical maximum energy efficiency.} 
We use a similar approach to prior work~\cite{energy-roofline-model} to
 obtain a roofline for energy efficiency. 
Figure~\ref{fig:roofline-energy} (right) shows the
 energy roofline for \am{Edge TPU}, along with the
efficiency achieved with each model.
 Note that unlike the throughput roofline model, the energy roofline
 is a smooth curve because we cannot hide memory energy (as opposed
 to memory transfer time, which can be overlapped with computation time and results in the sharp knee
for the throughput roofline). We find that across
 all models, the accelerator fails to get close to the maximum energy efficiency,
falling 62.4\% lower on average.
The energy efficiency is
 particularly low (33.8\% on average) for LSTM and Transducer models, but even the best CNN model
achieves only 50.7\% of the maximum efficiency.

\begin{figure}[h]
    \centering
        \centering
        \includegraphics[width=\linewidth]{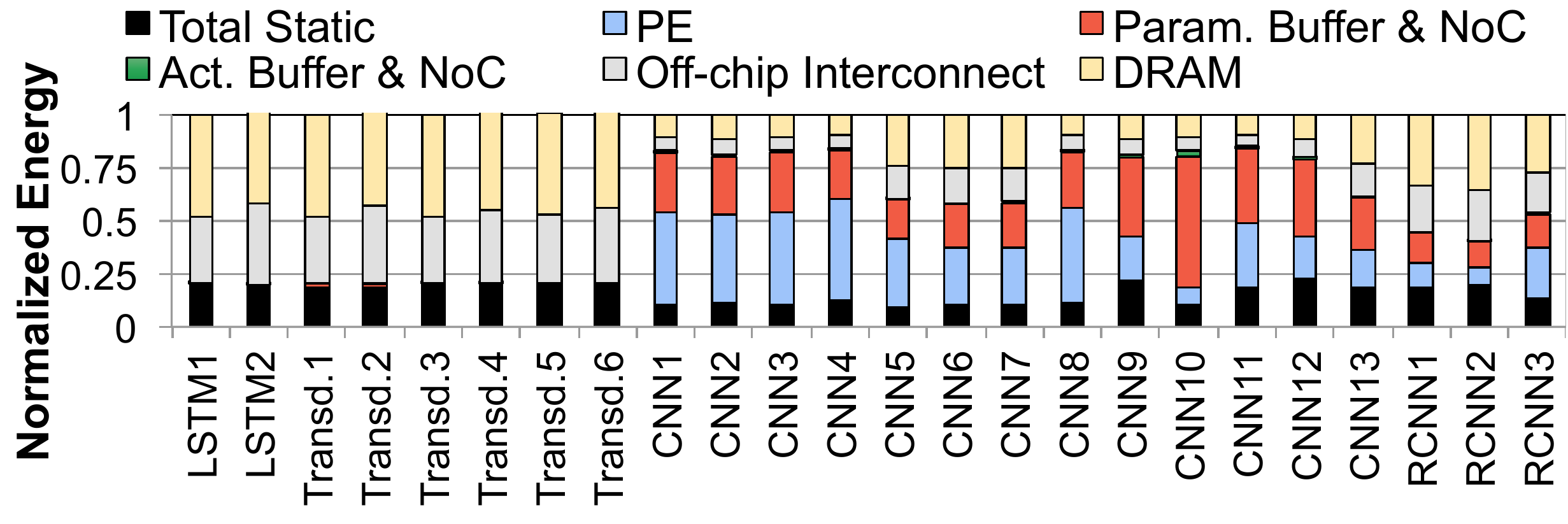}%
\vspace{-24pt}
    \caption{Energy breakdown during inference execution. \todob{2 models missing?}}
    \label{fig:energy-breakdown}
\end{figure}

\textbf{3. The accelerator's memory system design is neither effective nor efficient.}
Figure~\ref{fig:energy-breakdown} shows the energy breakdown during inference
 execution across different models. We make three key observations from this figure. 
 First, 
 on-chip buffers account for a significant portion of both static and dynamic energy, mostly because
 of their large size .
 For example, for CNN models, 48.1\%
 of the static energy and 36.5\% of the dynamic
 energy is spent on accessing and storing parameters in the on-chip buffers. Second, 
while the buffers consume a significant amount of area
(79.4\% of the total accelerator area) and energy,
they often do not reduce off-chip memory accesses.
For Transducers and LSTMs,
 only 11.9\% of the model parameters can fit into the buffer,
and for CNNs
the on-chip buffer can accommodate only 15.0\% of the parameters on average.  
Increasing the buffer size does not help:
if we increase the buffer size to 8x,
this provides a latency and energy reduction of only 37.6\% and 40.3\%, respectively,
for Transducers.
This is because
(1)~even such a large buffer can cache only 46.5\% of the parameters; and
(2)~the leakage and access energy increase significantly with a larger buffer size,
offsetting other benefits of a larger capacity.
Third, both on-chip (i.e., distributing parameters from the on-chip buffer to the PEs)
and off-chip (i.e., accesses to DRAM) network traffic for model parameters results in
high energy and performance costs.
On average, 50.3\% of the total \am{Edge TPU} energy is spent on off-chip
parameter traffic (DRAM accesses and off-chip interconnect),
and 30.9\% of the total energy is spent on on-chip parameter traffic.
We conclude that the \am{Edge TPU's} memory system
is highly inefficient.

\subsection{Layer-Level Study of Google Edge Models}
\label{sec:motiv:model}

To understand where \am{the Edge TPU's~\cite{edge-tpu}} pitfalls 
come from, we analyze the models themselves in significant detail.

\subsubsection{Analysis of LSTMs and Transducers}
\label{sec:motiv:model:lstm}

We identify three key properties of LSTMs and Transducers in our edge model analysis.

\begin{figure}[h]
    \vspace{-10pt}
    \centering
        \centering
        \includegraphics[width=\linewidth]{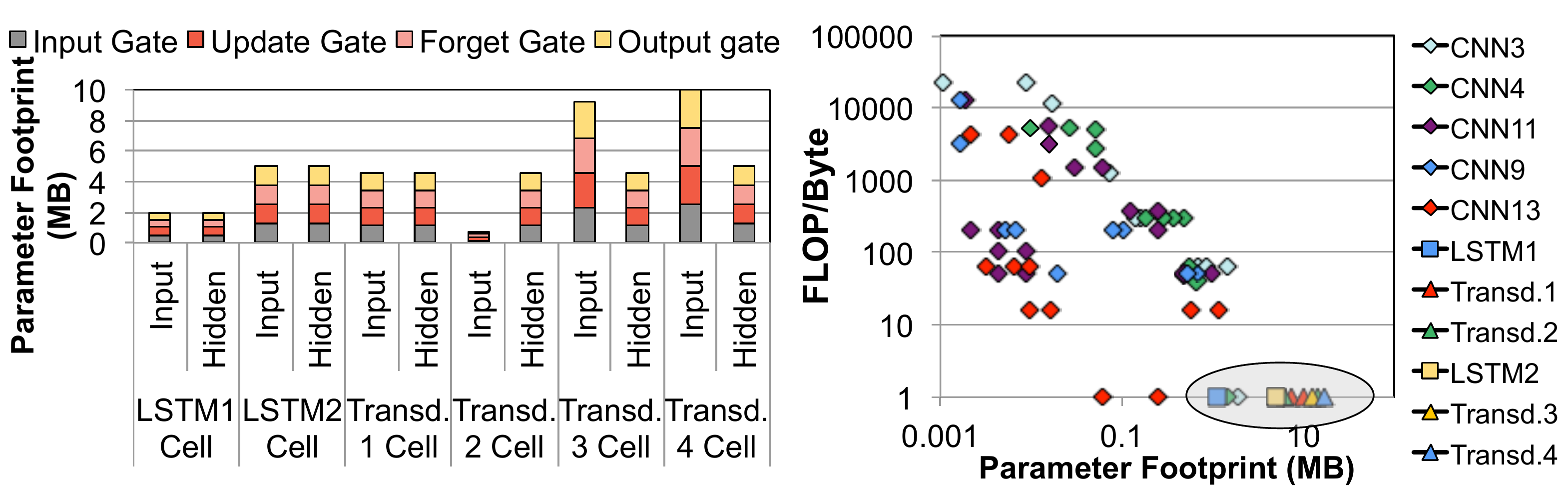}%
\vspace{-22pt}
    \caption{Parameter footprint of $W_{x}$ and $W_{h}$ for different LSTM gates for LSTMs and Transducers (left), and layer parameter footprint vs.\ 
  \si{\flop\per\byte} (right).}
    \label{fig:footprint-lstm-cell}
    \label{fig:param-reuse}
    \vspace{-8pt}
\end{figure}

\textbf{1. Large parameter footprint.}
Each gate in an LSTM cell has an average of 2.1 million parameters, which
includes parameters for both input ($W_{x}$) and hidden ($W_{h}$) matrices
(as shown in Figure~\ref{fig:footprint-lstm-cell}, left).
The large parameter footprint of LSTM gates results in large footprints for
LSTM layers (up to 70 million parameters), and in turn, LSTM and Transducer models
that include such layers.
Figure~\ref{fig:param-reuse} (right)
 shows the total footprint vs.\ the \si{\flop\per\byte} ratio (which indicates arithmetic intensity)
 across the layers of representative CNNs, LSTMs, and Transducers (the trend is the same across all models).
We observe from the figure that that layers from LSTMs and Transducers have significantly larger footprints 
(with an average footprint of \SI{33.4}{\mega\byte}) than layers from CNNs.

\textbf{2. No data reuse and low computational 
complexity.} For these layers, the \si{\flop\per\byte} of both $W_{x}$ and $W_{h}$ is one,
as shown in Figure~\ref{fig:param-reuse} (right).
The accelerator fetches $W_{x}$ and $W_{h}$ for each LSTM gate from DRAM, accesses them once to
 perform the input and hidden MVMs, and then does not touch the parameters again
 until the next LSTM cell computation, resulting in no reuse. 
Furthermore, these models have much a lower computational complexity than CNNs and RCNNs,
with 67\% fewer MAC operations on average.

\textbf{3. Intra- and inter-cell dependencies.} 
Two types of dependencies exist within LSTM layers, both of which
affect how the accelerator schedules gate computations:
(1)~\emph{inter-cell dependencies}, 
 which are due to recurrent connections ($h_{t}$) across cells; and 
(2)~\emph{intra-cell dependencies}, which
 enforce the following order between gates: all four gate computation need to be done
 before the cell state ($c_{t}$) gets updated, after which the hidden vector ($h_{t}$)) is generated. 
 To respect these dependencies, the accelerator 
 schedules cells computation in a
 sequential manner.
 This means that cell computation
 cannot start until $h_{t-1}$ is generated by the previous cell. 
 For each LSTM cell, the accelerator treats each gate as two
 fully-connected (FC) layers (corresponding to input MVM and hidden MVM),
 and runs the gates sequentially.

However, we find that this scheduling is not efficient. Due to the sequential execution of 
 FC layers, the latency of updating $c_{t}$ and generating $h_{t}$ become significantly
 higher, which in turn degrades PE utilization (as the PEs spend more time waiting for
 cell calculation to be completed). We find that, despite these dependencies, there
 are still multiple ways to parallelize computations within LSTM cells and improve
 PE utilization. For example, gates within each cell can be scheduled in any
 order, since there are no dependencies between the gates,
and the input and hidden MVMs within each gate are independent. 
 Since the scheduler in the baseline accelerator simply treats each gate computation as an FC layer (similar to the FC layer in
CNN models), it is oblivious of these opportunities for parallelization.

\subsubsection{Analysis of CNNs}
\label{sec:motiv:cnn}

\begin{figure}[h]
    \vspace{-10pt}
    \centering
        \centering
        \includegraphics[width=0.95\linewidth]{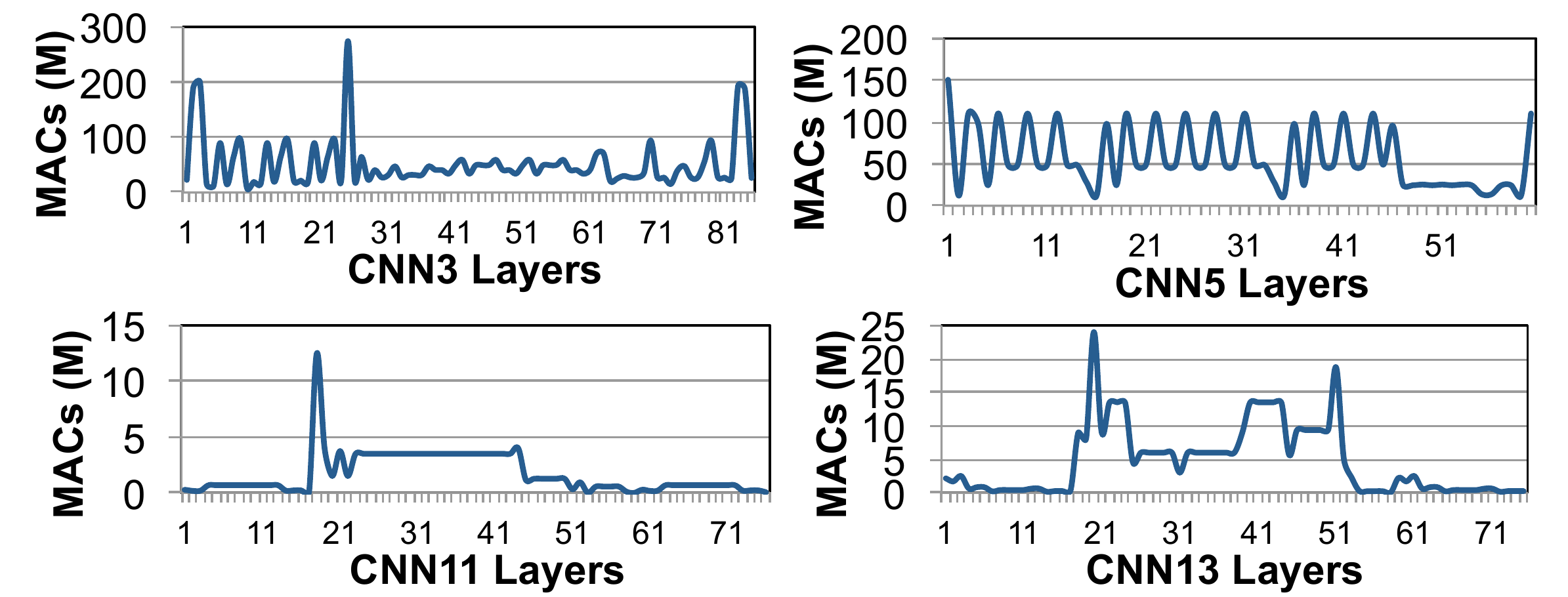}%
\vspace{-12pt}
    \caption{Analysis of number of MAC operation across different layers for four CNN models.}
    \label{fig:cnn-mac}
    \vspace{-14pt}
\end{figure}

\begin{figure}[h]
    \vspace{-5pt}
    \centering
        \centering
        \includegraphics[width=0.95\linewidth]{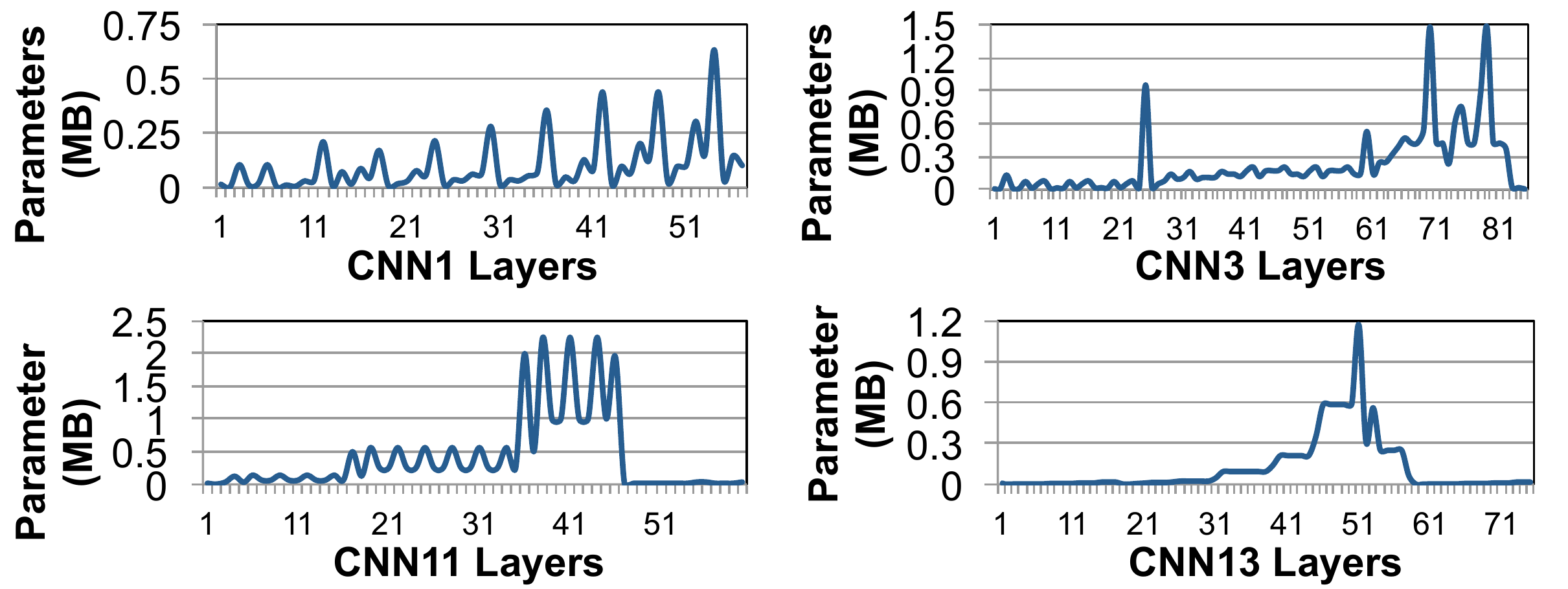}%
\vspace{-12pt}
    \caption{Analysis of parameter footprint across different layers for four CNN models.}
    \label{fig:cnn-param}
    \vspace{-8pt}
\end{figure}

Our analysis of edge CNN models reveals two interesting insights. 
First, 
we find that edge CNN
 models
unlike layers in traditional CNNs 
(e.g., AlexNet~\cite{alex-net}, VGG~\cite{simonyan2015very}), which tend to be
 relatively homogenous, we find that the layers in
 edge CNN models exhibit significant heterogeneity in terms of type
 (e.g., depthwise, pointwise), shape, and size. 
We find that this is often because these models
 employ several decomposition techniques~\cite{mobilenet,squeezenet} to reduce the computational
 complexity and footprint of layers, in order to make them more friendly for the
 constrained edge devices. 
As an example of layer diversity,
Figures \ref{fig:cnn-mac} and \ref{fig:cnn-param} show the number of MAC operations and 
parameter footprint across different layers for four CNN models. 
We find that the MAC intensity and parameter footprint
 vary by a factor of 200x and 20x across different layers.

Second, we find that layers exhibit significant variation in terms of data reuse patterns for
 parameters and for input/output activations. For example, a depthwise layer includes only
 one channel, which leads to very low activation reuse, while a pointwise layer takes in a
$1 \times K$ filter (\emph{K} is the depth of the input channel) and convolves it with
an input activation across different input channels, leading to
much higher activation reuse.
We also observe variation in data reuse across layers of the same type.
For example, initial/early standard convolution layers
in edge CNNs have a shallow input/output channel depth, large input activation
width/height, and very small kernels, resulting in very high
parameter reuse. In comparison, standard convolution
layers that are placed toward the end of the network have
a deep input and output channel depth, small input activation
width/height, and a large number of kernels, resulting in very low parameter reuse.  
This variation in reuse is illustrated in Figure~\ref{fig:param-reuse} (right),
 which shows that the \si{\flop\per\byte} ratio varies
across different layers for five representative CNN models
by a factor of 244x.

\subsubsection{Analysis of RCNNs}

RCNNs include layers from both CNNs and LSTMs. As a result, individual layers from RCNN models
exhibit the same characteristics we discussed above for LSTM and CNN layers. 
We find that layers from RCNNs exhibit on average significantly higher footprints and lower \si{\flop\per\byte}
ratios than CNNs, as they include both LSTM and CNN layers.
Due to the inclusion of both layer types, we observe significantly more variation across
RCNN layers characteristics as well.

\subsubsection{Sources of Accelerator Pitfalls}
\label{sec:motiv:discussion}

\paratitleattop{}
\paratitle{PE Underutilization} 
We identify three
 reasons why \am{the Edge TPU} falls significantly short of peak throughput.
First, while some layers have high parameter reuse
(e.g., pointwise layers, with a \SI{1200}{\flop\per\byte} ratio), 
other layers exhibit very low reuse (\SIrange{1}{64}{\flop\per\byte})
while having large parameter footprints (\SIrange{0.5}{5}{\mega\byte}).
\sg{Layers with low reuse yet large footprints for parameters
often leave PEs idle, as the parameters incur cache misses to DRAM.}
The bandwidth of modern commercial DRAM
(e.g., \SI{32}{\giga\byte\per\second} for LPDDR4~\cite{lpddr4}) is two orders 
of magnitude below the \SI{2}{\tera\byte\per\second} bandwidth needed 
to sustain peak PE throughput when only one \si{\flop\per\byte} is performed
\sg{(as is the case for LSTM layers)}.
Many (but not all) layers end up with a 
similar bandwidth problem.

Second, the accelerator does not provide a custom dataflow optimized for each layer.
As we identified in Section~\ref{sec:motiv:model}, layers both across and within models
exhibit high variation in terms of data reuse patterns.
This variation necessitates the need for \emph{different} dataflows for different layers,
where each data flow exposes a different set of reuse opportunities for
parameters and activations.
However, 
\sg{state-of-the-art accelerators such as the Edge TPU employ} a \emph{single}
dataflow that is designed for high spatial/temporal reuse~\cite{maeri, meastro}.
The missed reuse opportunities in many of the model layers causes PEs
to needlessly wait on retrieving previously-accessed data that was not properly 
retained on-chip.

Third, the different shapes and inter-/intra-layer dependencies across 
different types of layers (e.g., LSTM cell, standard convolution, depthwise, 
pointwise, fully-connected)
makes it challenging to fully utilize a PE array with a fixed size,
which is the case in state-of-the-art accelerators~\cite{eyerissv2}.
To cater to these differences across layers, there is a need for both
better scheduling (e.g., uncovering parallel computation opportunities
as we found in Section~\ref{sec:motiv:model:lstm})
and appropriately sizing the PE arrays
in order to maintain efficient utilization.

\paratitle{Poor Energy Efficiency}
We find three major sources of energy inefficiencies.
 First, the accelerator incurs high static energy costs
 because (1)~it employs a large overprovisioned on-chip buffer,
 and (2)~it underutilizes PEs. 
  Second, 
the on-chip buffers consume a high amount of dynamic energy, 
as we saw for CNN layers in Section~\ref{sec:motiv:accelerator}.
Third, the accelerator suffers from the high cost 
of off-chip parameter traffic.
On-chip buffers
 fail to effectively cache parameters for many layers due to layer diversity, causing
 50.3\% of the total inference energy to be spent on
 off-chip parameter traffic.

\paratitle{Memory System Issues}
We uncover two sources of memory system challenges.
First, due to layer diversity, 
on-chip buffers are ineffective for a large fraction of layers. 
As we discuss in Section~\ref{sec:motiv:model:lstm},
LSTM gates have large parameter footprints and 
zero parameter reuse,
rendering the on-chip buffer useless for a majority of
LSTMs and Transducers, and for a significant fraction of RCNN layers.
For CNN layers, we
find that those layers with low data reuse account for a significant portion
 of the entire model parameters (e.g., 64\% for CNN6). This means that the on-chip buffer fails
to cache a large portion of the parameters for CNN models.
As a result, despite being several megabytes in size,
the on-chip buffer is effective only for a small fraction of layers,
which have an average parameter footprint of only \SI{0.21}{\mega\byte}.

Second, due to having an
unnecessarily large on-chip parameter buffer, accesses from
those layers with high data reuse become very costly. For
those layers, even though they have a small footprint, they perform
a large number of buffer accesses.
As the on-chip buffer is overprovisioned,
the large capacity leads to a high dynamic energy cost per access,
even though the capacity is not needed.

\subsection{Key Takeaways}

\am{Our analysis provides three key insights:
(1)~there is significant variation in terms of layer characteristics \emph{across} and 
\emph{within} state-of-the-art Google edge models, (2)~the monolithic design of the Edge TPU
is the root cause of its shortcomings, and (3)~to achieve high utilization and energy efficiency, 
all key components of an edge accelerator (PE array, dataflow, on-chip memory,
off-chip memory bandwidth)
must be customized based on layer characteristics. 
}


\section{\texorpdfstring{\MakeUppercase\titleShort}{\titleShort} Framework: Key Ideas}
\label{sec:proposal}
\label{sec:framework}

\titleShort is a new \sg{acceleration} framework 
that \sg{harnesses layer variation across edge NN models for} high efficiency.

\subsection{High-Level Overview}
\label{sec:framework:overview}

\titleShort distributes the layers from an NN model across
a collection of smaller hardware accelerators that are carefully specialized towards the
properties of different layer types.
A drawback of current monolithic accelerators is that by designing
for a wide range of layers, many of their hardware resources are overprovisioned
or introduce high inefficiency.
By specializing each accelerator to a subset of layers, \titleShort avoids 
these issues, resulting in an accelerator with a much smaller area that 
achieves high throughput and efficiency for those layers.
\titleShort consists of (1)~a collection of heterogeneous hardware accelerators;
and
(2)~a runtime scheduler that determines which accelerator each layer
in a model should execute on, using a combination of model and hardware
characteristics.
As we show in Section~\ref{sec:arch}, layers tend to group together into
a small number of clusters, allowing designers to typically keep the number of
accelerators low.

\titleShort is designed as a framework that can support a wide range of
architectural implementations.  This allows \titleShort to be optimized to
specific system needs, which is critical to keep resource utilization to a
minimum in resource-constrained edge devices, and lets the framework adapt
easily to future types of NN models that we expect will arise in the future.
We discuss one example implementation of \titleShort in Section~\ref{sec:arch},
which caters to the Google edge models that we analyze,
to illustrate the effectiveness of our framework.

\subsection{Scheduler}
\label{sec:proposal:scheduler}
\label{sec:framework:scheduler}

The goal of {\titleShort}'s software runtime scheduler is to identify 
which accelerator each layer in an NN model should run on.
Each of the accelerators in \titleShort caters to a specific cluster of layers,
where the cluster is defined using specific layer characteristics
(e.g., type, footprint, data reuse, dependencies).
For a given hardware configuration, the scheduler has two pieces of 
information:
(1)~the characteristics of each cluster; and
(2)~which hardware accelerator is best suited for each cluster.
Similar to how chipset drivers are configured, this information is generated
once during initial setup of a system, and can be modified with an updated
driver version to account for new clusters.

When an NN model runs on \titleShort, the scheduler generates a mapping 
between layers and accelerators in \titleShort.
The scheduler uses the NN model (including a directed acyclic graph that 
represents communication across model layers) and the configuration information
in the driver to determine this mapping.
The mapping is generated in two phases. 

In the first phase, the scheduler iterates through each layer in the model, and 
identifies the ideal hardware accelerator for each layer \emph{in isolation}
(i.e., without considering communication overhead).
The scheduler determines two properties for each layer: 
(1)~the cluster that the layer belongs to; and
(2)~the target accelerator for the layer; 
While this phase works to maximize accelerator throughput and energy efficiency for each layer,
the resulting schedule may be sub-optimal for efficiency, because it does not
consider the overhead of transferring activations or communicating 
dependencies (e.g., $h_{t}$ in LSTM cells)
between different layers.
This can have a large impact on the overall framework performance and energy if
the amount of communication is large.

In the second phase, our scheduler accounts for the communication
overhead using a simple cost analysis algorithm.
For each layer, if the following layer (which we call the \emph{subsequent layer})
is mapped by the first phase to a different hardware accelerator than the 
current layer, the algorithm calculates the total amount of
data that needs to be communicated between the two layers, and uses a
simple heuristic 
to estimate the total cost 
of 
(1)~using the phase one mapping
(i.e., moving data to the subsequent layer's accelerator and running that layer
on its optimal accelerator), or
(2)~moving the subsequent layer
(i.e., running the subsequent layer on the current layer's accelerator, 
which eliminates accelerator-to-accelerator communication, but introduces 
inefficiencies due to running the subsequent layer on a sub-optimal accelerator).
If the cost of moving the subsequent layer is lower than the cost of
using the phase one mapping, the subsequent layer is remapped to the
current layer's accelerator.

Once the second phase is complete, \titleShort begins model execution
using the generated mapping.
\am{The scheduler ensures that each layer is executed completely in a single accelerator. Thus, there is no need
 for intra-layer data synchronization. On the other hand, due to the dependencies between layers, different accelerators
 may need to communicate with each other. Parameters are read-only and are always fetched
 from DRAM (i.e., no need for data synchronization). However, activations need to
 be transferred if two consecutive layers run on different accelerators. \titleShort uses DRAM to synchronize 
activations across accelerators, which avoids the need for sophisticated coherence mechanisms between on-chip
 and near-data accelerators~\cite{conda}.}


\section{\texorpdfstring{\MakeUppercase\titleShort}{\titleShort} for Google Edge Models}
\label{sec:arch}

We now discuss \sg{an example \titleShort design optimized} for our Google edge
models.
We start by identifying layer clustering in these models
(Section~\ref{sec:arch:clusters}).
Using the unique characteristics for each cluster, we determine which
characteristics have the greatest impact on accelerator design, and use
that to guide the number of accelerators that we need
(Section~\ref{sec:arch:hw}).

\subsection{Identifying Layer Clusters}
\label{sec:proposal:analysis}
\label{sec:arch:clusters}

\begin{figure}[h]
    \centering
        \centering
        \includegraphics[width=\linewidth]{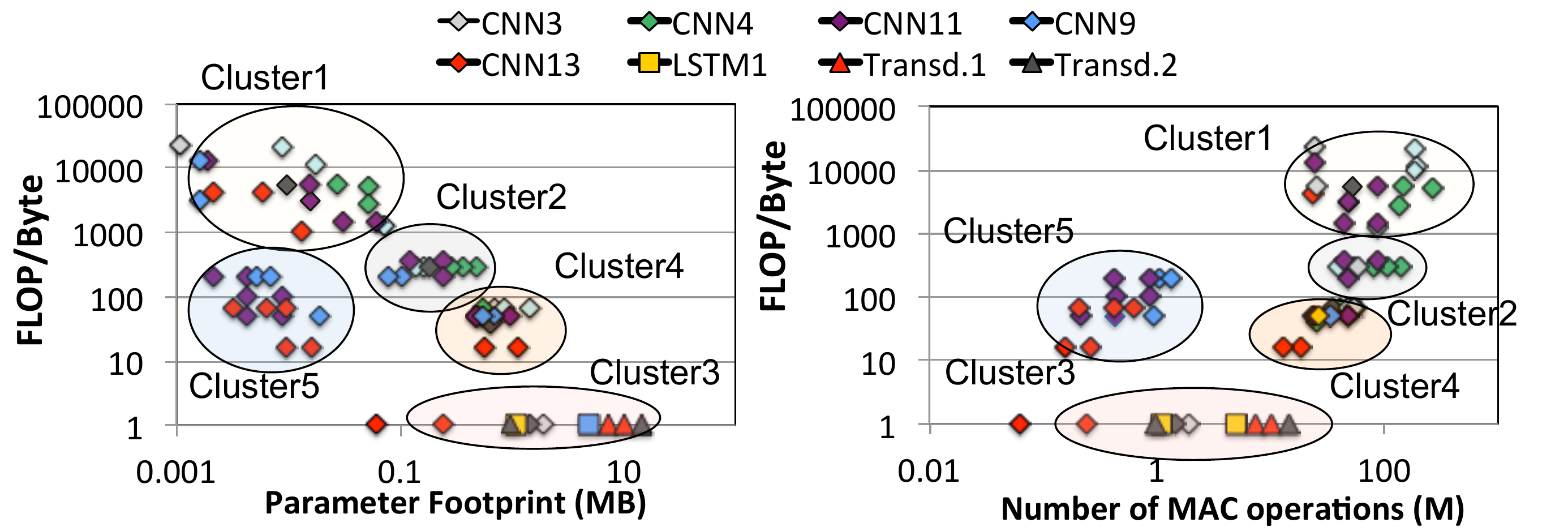}%
\vspace{-24pt}
    \caption{Parameter footprint vs. parameter \si{\flop\per\byte} ratio (left) and number of MAC operations (in millions) vs. parameter \si{\flop\per\byte} ratio (right) across layers.}
    \label{fig:layer-analysis}
\end{figure}

We revisit the models that we analyze in Section~\ref{sec:motiv}.
For each layer, we study the correlation between different characteristics.
As an example, Figure~\ref{fig:layer-analysis} shows
how the parameter footprint and the 
number of MAC operations correlate to parameter reuse
(\si{\flop\per\byte})
for a representative set of layers from 
five CNNs, two LSTMs, and two Transducers,
in order to improve figure clarity.
Based on all of the layer characteristics that we analyze,
we observe across all layers from all models (not just the representative
layers or correlations plotted) that 97\% of the layers group into one of
five clusters.

\paratitle{Cluster~1} 
Layers in this cluster have 
(a)~a very small parameter footprint (\SIrange{1}{100}{\kilo\byte}),
(b)~a very high \si{\flop\per\byte} ratio (780--20K), and
(c)~high MAC intensity (30M--200M). 
These layers exhibit high activation 
footprints and data reuse as well.
A majority of layers in this cluster are standard convolution 
layers with shallow input/output channels and large input activation width/height.
We find that these layers are mostly found among early layers in models,
and typically achieve high PE utilization (on average 82\%). 

\paratitle{Cluster~2}
Layers in this cluster have 
(a)~a small parameter footprint (\SIrange{100}{500}{\kilo\byte};
12x higher on average vs.\ Cluster~1),
(b)~a moderate \si{\flop\per\byte} ratio (81--400;
up to 10x lower than Cluster~1), and 
(c)~high MAC intensity (20M--100M).
They exhibit high activation footprints
and activation data reuse as well.
Many of the layers belong to pointwise layers, which have high parameter reuse
due to convolving 1x\emph{K} filters (\emph{K} is input channel depth) with input activation across different channels.
Other layers in the cluster include standard convolution layers commonly
found in the middle of CNN networks, with 
deeper input/output channels and smaller activation width/height than the 
convolution layers in Cluster~1, 
We find that Cluster~2 layers have lower PE utilization (64\%)
than layers in Cluster~1.

\paratitle{Cluster~3} 
Layers in this cluster have 
(a)~a very large parameter footprint (\SIrange{0.9}{18}{\mega\byte}),
(b)~minimal parameter reuse, and
(c)~low MAC intensity (0.1M--10M).
These layers exhibit small activation footprints but rather high activation reuse. 
The majority of these layers are from LSTM gates in LSTMs and Transducers,
or are fully-connected layers from CNNs.
These layers have very low PE utilization (0.3\% on average).

\paratitle{Cluster~4}
Layers in this cluster have 
(a)~a larger parameter footprint (\SIrange{0.5}{2.5}{\mega\byte}),
(b)~low-to-moderate \si{\flop\per\byte} ratio for parameters (25--64), and
(c)~moderate MAC intensity (5M--25M).
The layers exhibit small activation footprints but rather high activation reuse.
A large portion of layers in this category are standard convolution layers with
deep input/output channels and input activation width/height,
along with a large number of kernels. We find that these layers not only have large parameter footprint, but 
also include the majority of parameters ({up to 64.3\%) for their respective model.
Cluster~4 layers average a PE utilization of 32\%.

\paratitle{Cluster~5} 
Layers in this cluster have 
(a)~a very small parameter footprint (\SIrange{1}{100}{\kilo\byte}),
(b)~a moderate \si{\flop\per\byte} ratio for parameters (49--600), and
(c)~low MAC intensity (0.5M--5M).
The layers exhibit rather high activation footprints but have almost zero activation data
reuse. 
Many of these layers are depthwise layers, which have only one channel
(hence the lack of activation reuse), but have high parameter reuse because
a small number of filters are convolved across a large input activation
(which is the opposite of Cluster~4 layers)
Cluster~5 layers achieve a low average PE utilization of 21\%.

\subsection{Hardware Design Principles}
\label{sec:proposal:implication}
\label{sec:arch:hw}

As we study the distinguishing characteristics of each \sg{of our five clusters}, we find that
some characteristics have a strong influence on the hardware design,
while others do not necessitate
significant changes to the
hardware.
We discuss two insights that drive our hardware design decisions.

First, we find that MAC intensity and parameter footprint/reuse drive a
significant divergence in hardware design, as they impact a number of
key accelerator design parameters (e.g., PE array size,
on-chip buffer size, memory bandwidth considerations).
Looking at our five clusters, we identify that
(a)~layers in Clusters 1 and 2 share a high MAC intensity, 
smaller parameter footprint, and moderate-to-high parameter reuse; while
(b)~layers in Clusters 3 and 4 share a lower MAC intensity,
larger parameter footprint, and low parameter reuse.
This means that we need \emph{at least} two different accelerator
designs:
one that caters to the more compute-centric behavior of Clusters~1/2,
and one that caters to the more data-centric behavior of Clusters~3/4.
Given our resource-constrained environment, we look to see if layers in
Cluster~5, which have a low MAC intensity (similar to Clusters 3 and 4) but
a smaller parameter footprint (similar to Clusters 1 and 2),
can benefit from one of these two approaches.
We find that the low MAC intensity, along with the lower parameter reuse
by many Cluster~5 layers, allow the layers to benefit from many of the
non-compute-centric optimizations that benefit Clusters 3 and 4, so we
consider them together.

Second, 
\sg{a key distinguishing factor between different accelerator designs
is the accelerator dataflow, as it dictates which
reuse opportunities in layers are exploited,
and thus strongly impacts PE utilization and energy efficiency.}
Prior work~\cite{meastro} analyzes the large dataflow design space,
and discusses four types of data reuse:
\emph{spatial multicasting} (reading a parameter once, and spatially distribute
it as an input to multiple PEs),
\emph{temporal multicasting} (replicating a parameter in a small local buffer,
and delivering the parameter as multiple inputs at different times to the same PE),
\emph{spatial reduction} (accumulating activations from multiple PEs 
at the same time using multiple compute units), and
\emph{temporal reduction} (accumulating multiple activations generated at 
different times using a single accumulator/buffer).
The chosen dataflow
directly affects how the memory system and on-chip network of an accelerator
should be designed.
Thus, 
we need different dataflows for clusters with
significantly different reuse patterns.

Both of our compute-centric clusters
\sg{benefit from a similar}
dataflow, which exposes reuse opportunities for both parameters and
activations.
Between the compute-centric optimizations and the shared dataflow affinity,
we determine that we can use a single accelerator 
(\emph{\accelA})
to efficiently execute layers from both Cluster~1 and Cluster~2.

Across our three data-centric clusters, we find that
\sg{layers from both Clusters 4 and 5}
benefit from a dataflow that exposes
reuse opportunities for parameters
but not for activations,
and can thus share an accelerator
(\emph{\accelC}).
 Cluster~3 layers exhibit different data reuse characteristics,
\sg{instead benefitting}
from a dataflow that provides temporal reduction
 opportunities for activations.
As a result, \sg{they} need a separate accelerator
(\emph{\accelB}).

We employ a \emph{template-based} design approach:
while we design each accelerator based on cluster characteristics, we maintain the
same generic tiled architecture as the baseline accelerator. 
We do this to ease the integration of our hardware into a real system:
from the perspective of compilers and programs, each of our accelerators
appears to be just a different instance (with a different configuration) of
the baseline accelerator.
This is a \emph{critical design decision}, as it allows us to
 deploy and run models using existing highly-optimized design/compile toolchains
(e.g., Edge TPU compiler~\cite{edge-tpu-compiler}) 
seamlessly on all of \titleShort's accelerators, with the trade-off
of limiting the degree of customization (and, thus, efficiency) our
accelerator designs can achieve.
While \titleShort can work with accelerators that do not employ this
tiled architecture, we make this design choice to
avoid the need for more complex compilers and multiple libraries
for programmers.

\subsection{\accelA Accelerator Design}
\label{sec:arch:comp-x}

\paratitleattop{}
\paratitle{Dataflow}
\label{sec:arch:comp-x:dataflow}
\accelA caters to 
Cluster~1/2 layers, which have opportunities for 
both parameter and activation reuse.
The layers have a small parameter footprint,
which can be captured by a very small on-chip buffer.
As a result, exposing reuse opportunities for parameters will likely not
provide much additional benefit.
The large output activation footprint would require a relatively large buffer
to effectively exploit reuse, which would consume significant area and 
static energy.

\sg{If we design our dataflow to provide temporal reduction for the
activations, we can significantly reduce the required buffer size
while still enabling high reuse (\emph{first requirement}).
However, not all such dataflows can improve efficiency.
One example is a dataflow where activations are spatially distributed
across PEs, allowing the PEs to collaboratively calculate each output.
We should how a pointwise layer (Figure~\ref{fig:compx-dataflow}a)
would map to such a dataflow in Figure~\ref{fig:compx-dataflow}b.
Unfortunately, this dataflow actually \emph{worsens} the impact of the
large output activation footprint, as each output activation is now split into
partial sums (one per PE) that need to be gathered and reduced.
This reduction generates significant traffic on the on-chip network.
We instead want 
a dataflow that \emph{avoids} spatial reduction 
for output activations (\emph{second requirement}).}

\begin{figure}[h]
    \centering
        \includegraphics[width=0.75\columnwidth]{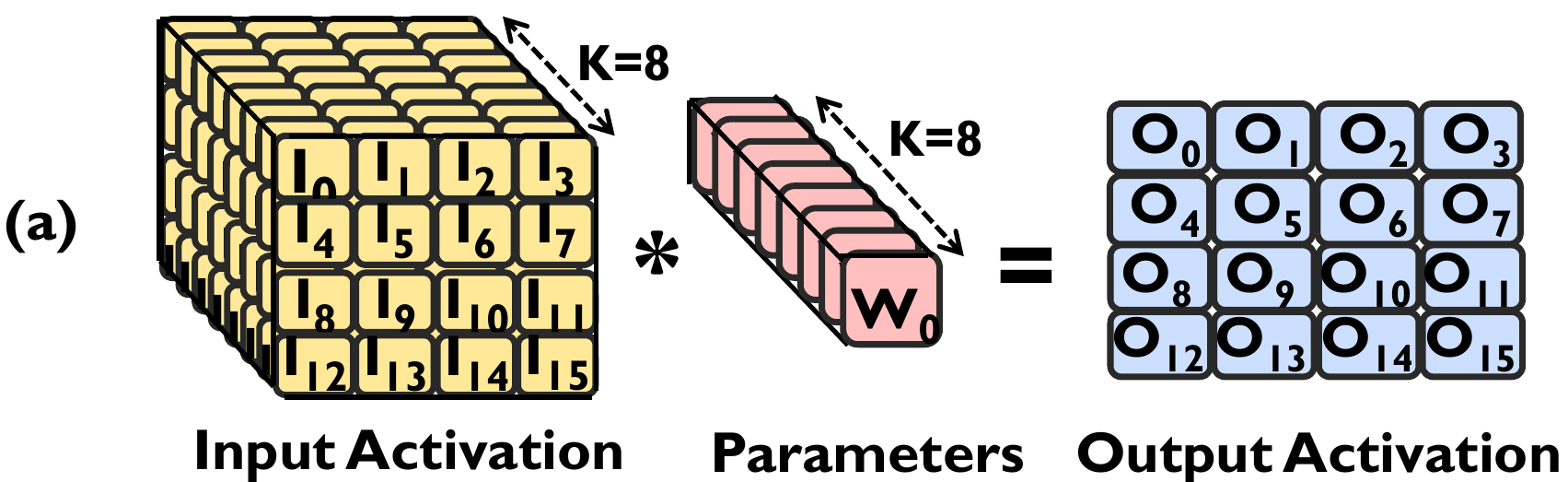}\vspace{2pt}\\%
        \includegraphics[width=\columnwidth]{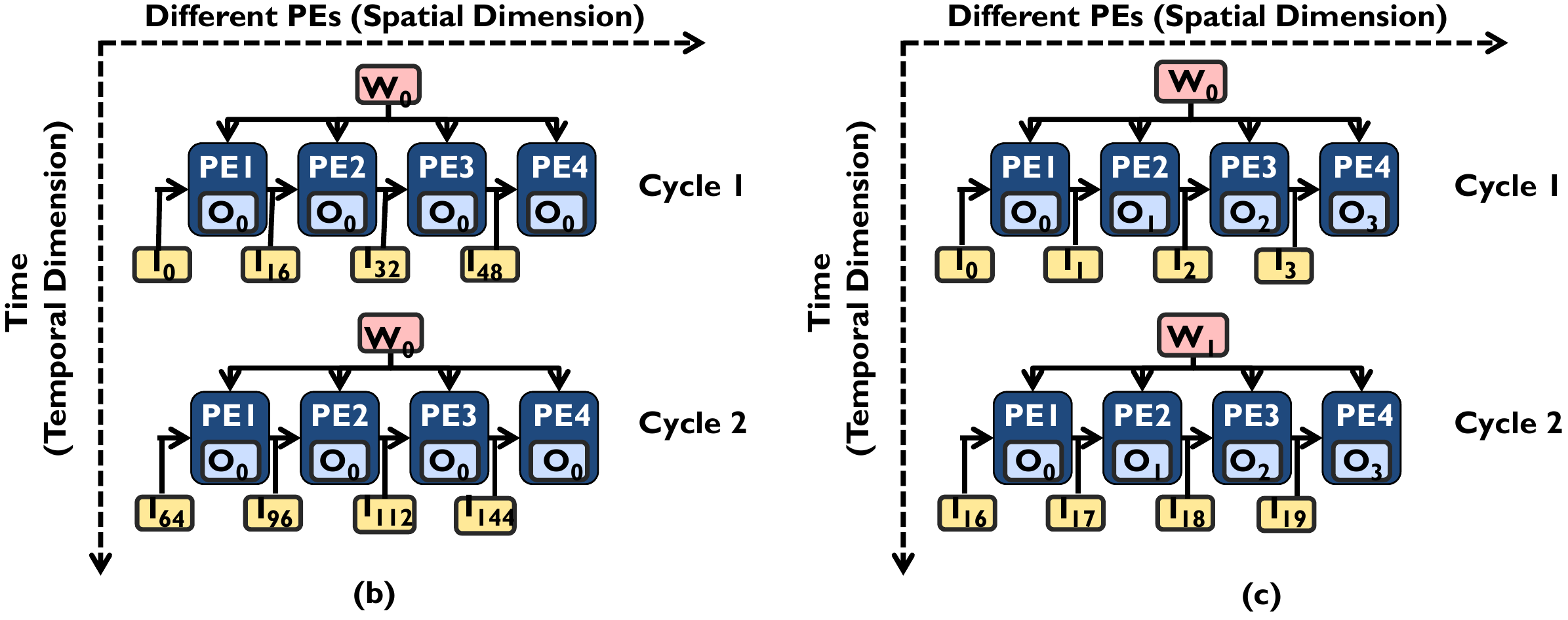}%
\vspace{-24pt}
    \caption{(a)~Simplified pointwise layer, (b)~an example dataflow that generates high network traffic by spatially reducing output activations, (c)~our proposed dataflow.}
    \label{fig:compx-dataflow}
\end{figure}

\sg{We develop a dataflow for \accelA that meets both
of our design requirements, as shown using a 4-PE toy example
in Figure~\ref{fig:compx-dataflow}c.}
We spatially distribute output
activations, such that each PE calculates one output activation
element. This enables temporal reduction for each output activation,  
 and reduces the on-chip buffer size.
We replicate each parameter across PEs
at each time step (i.e., each PE receives the same parameter),
while spatially distributing input activations across PEs.
This provides spatial reuse (multicast) for parameters across all PEs in each cycle.
Our proposed dataflow generates no partial sum traffic,
as the partial sums are instead \emph{temporally reduced}
in each PE's register file.

\paratitle{PE Array}
\label{sec:arch:comp-x:pe}
Layers from Clusters 1 and 2 perform a large number of MAC operations, so
we want to size \accelA's PE array to efficiently perform these in parallel.
We analyze the inference latency using different array configurations, and
empirically find that a 32x32 array strikes a balance between latency,
PE utilization, and energy consumption.
In particular, a smaller array increases the latency for Cluster~1 layers,
while a larger array provides few benefits.
With this array, \accelA achieves a \SI{2}{\tera\flop\per\second} peak throughput.

\paratitle{Memory System}
\label{sec:arch:comp-x:mem}
The data reuse opportunities exposed by
 our proposed dataflow enable us to significantly reduce
 the on-chip buffer sizes, significantly reducing the energy and area
 costs of \accelA. 
 As we mentioned above, by enabling temporal reduction (in output activations) with our dataflow,
we can reduce the size of the on-chip buffer that holds the activations
from \SI{2}{\mega\byte} down to \SI{256}{\kilo\byte}.
Thanks to the small parameter footprint of layers in Clusters 1 and 2,
we can reduce the on-chip parameter buffer by 32x, down to \SI{128}{\kilo\byte}.

\subsection{\accelB Accelerator Design}
\label{sec:arch:mem-x}
\label{sec:arch:mem1-x}

\paratitleattop{}
\paratitle{Dataflow}
\label{sec:arch:mem-x:dataflow}
\label{sec:arch:mem1-x:dataflow}
We leverage three characteristics of Cluster~3 layers to
shape the design of \accelB: 
(1)~parameters exhibit zero reuse but have a very large
 footprint, 
 (2)~activations exhibit high reuse but have a small footprint, and
 (3)~the main operation is MVM.
Based on these, we identify two design requirements for our dataflow.

First, our dataflow should expose activation reuse opportunities during MVM.
One way is using spatial reduction for output activations.
However, due to the large parameter footprint (e.g., $W_{h}$ in LSTM cells),
this would experience similar issues as we saw for spatial reduction
for Clusters 1 and 2 (Section~\ref{sec:arch:comp-x:dataflow}):
there would be a large number of partial sums that need to be
combined (as many as one per PE every cycle), resulting 
in significant network traffic.
As a result, we look for more efficient ways to expose activation reuse.

Second, we need to effectively utilize memory bandwidth at low cost.
The on-chip buffer is ineffective for these layers, due to the lack of data reuse and the
large parameter footprint.
As a result, high PE utilization requires many off-chip accesses, and, thus, 
access to high bandwidth, such as the bandwidth
available \emph{inside} modern 3D-stacked memories.
While we want to fully utilize this bandwidth, we cannot do so simply by
issuing many outstanding memory requests at once from the accelerator, 
as this requires complex hardware that often exceeds the
area and power budgets available in these memories 
(and in edge devices)~\cite{tetris, mondrian, google-pim}.
However, if we can design our dataflow to issue \emph{sequential}
memory accesses to parameters, we can exploit this sequential
pattern to use the bandwidth without complex hardware and at a much lower
energy cost~\cite{tetris, mondrian}.

Using these requirements, we design our \accelB dataflow.
Output activations are spatially distributed such that each PE is responsible for
 one output element. Every cycle, the dataflow replicates each input element
across all PEs. 
We use spatial parameter distribution, with each PE receiving a different
parameter element during a cycle.
Figure~\ref{fig:lstm-non-opt}a shows a toy example of MVM
 for the forget gate in a LSTM cell, 
while Figure~\ref{fig:lstm-non-opt}b shows an example of how \accelB's dataflow
runs that MVM operation on an accelerator with 4 PEs.  
Each cycle, all of the PEs multiply one input vector element
($X_{0}(t)$ in this example) with a sub-row of $W_{x}$ (or $W_{h}$),
storing the partial sum in the PE register file.
In the next cycle, the next input
element ($X_{1}(t)$) is replicated across all PEs, the next sub-row of parameters is 
spatially distributed, and the multiply is performed.
The product is added to the partial sum sitting in the PE register file
\sg{(which enables temporal reduction for output activations)}.
After we multiply all rows of the parameter matrix
with input vector elements, we obtain the final output activation in each PE.

\begin{figure}[h]
    \centering
        \centering
        \includegraphics[width=\linewidth]{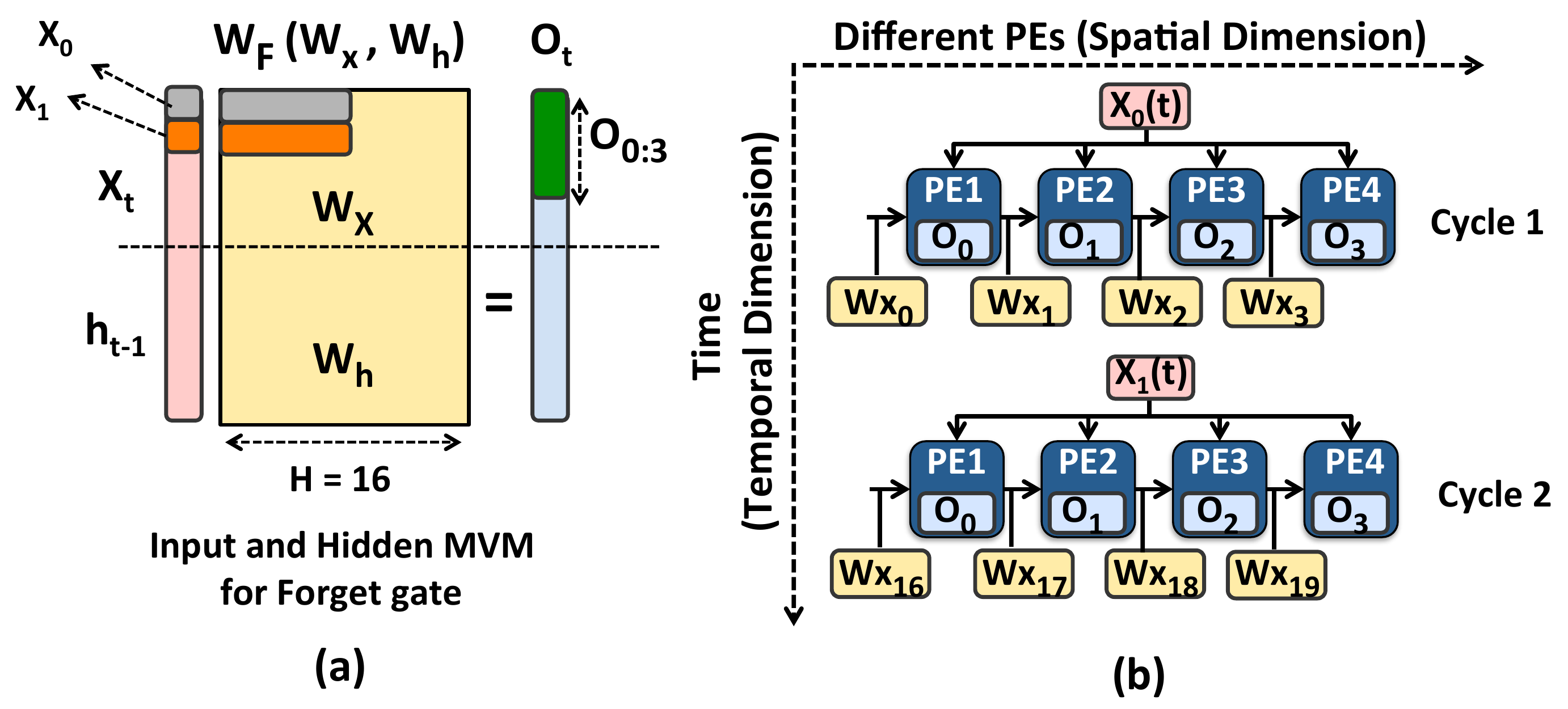}%
\vspace{-26pt}
    \caption{(a)~Input/hidden MVMs for a forget gate in a LSTM cell, (b)~our proposed dataflow for
        an LSTM cell.}
    \label{fig:lstm-non-opt}
\end{figure}

\sg{While the above dataflow is efficient for a single MVM operation,
it does not account for inter- and intra-cell dependencies in LSTM layers
(the majority of layers in Cluster~3).
Not exploiting reuse opportunities for these dependencies can
be costly (e.g., for a 100-sequence input, where
$W_{x}$ and $W{h}$ are both 1000x1000,
each layer would generate \SI{500}{\mega\byte} of
off-chip parameter traffic).
$W_{x}$ and $W_{h}$ elements are reused 
across LSTM cells within a layer (e.g., $W_{hf}$
is reused across all cells in Figure~\ref{fig:lstm-opt}a).
While these arrays are too large to cache,
we can decouple the computation of input and hidden MVM
in each gate, and calculate all input MVMs for \emph{all} cells
ahead of time (as they have no dependencies across cells).
This allows us to fetch $W_{hf}$ elements only once from memory.
We can later perform the hidden MVM computations
while respecting their inter-cell dependencies.}

\begin{figure}[h]
       \centering
        \includegraphics[width=\linewidth]{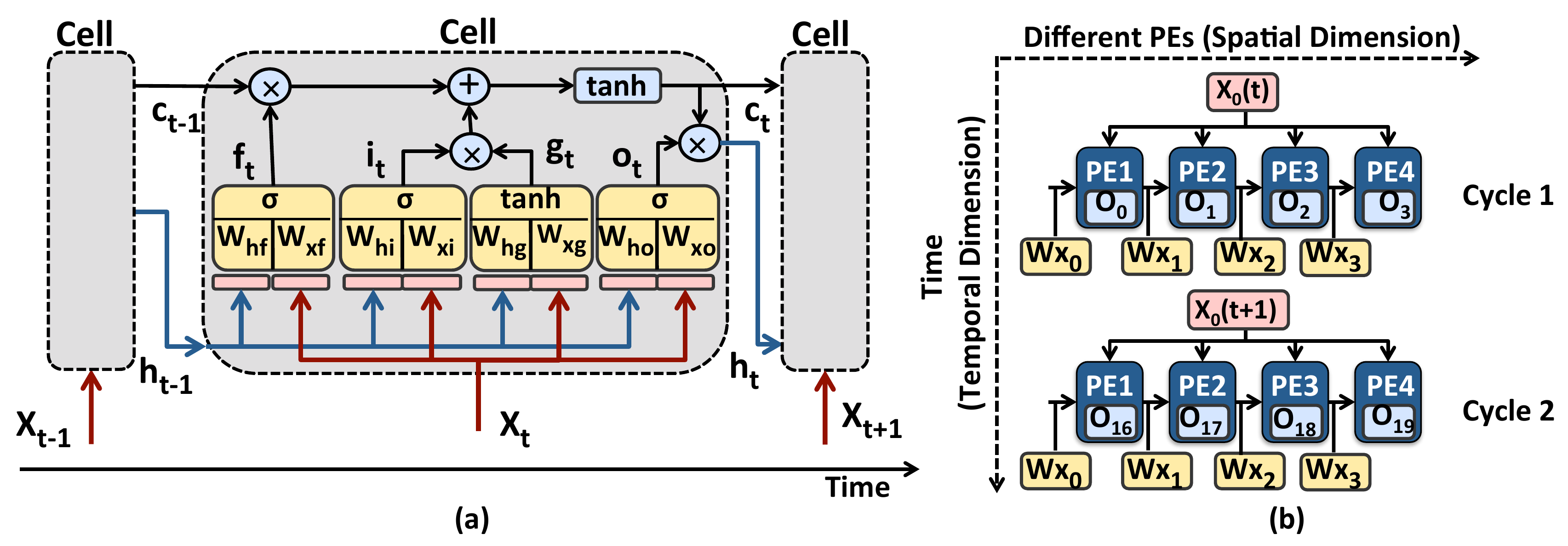}%
\vspace{-24pt}
    \caption{ (a)~LSTM layer internal structure, (b)~our LSTM cell dataflow optimized to exploit dependencies.}
    \label{fig:lstm-opt}
\end{figure}

We modify our dataflow, as shown in Figure~\ref{fig:lstm-opt}b, to
efficiently reuse $W_{x}$ and $W_{h}$.
Now, when we fetch and spatially distribute a sub-row of $W_{x}$
across PEs,
we keep those parameters in the PE register file.
Each iteration, every PE multiplies one element of $x_{t}$ ($x_{0}(t)$ in Figure~\ref{fig:lstm-opt}b)
and one element
of the $W_{x}$ sub-row.
Now, instead of going to the next sub-row, we fetch the next input element
($x_{0}(t+1)$ in Figure~\ref{fig:lstm-opt}b) and perform the MVM
for another LSTM cell.
Note that instead of accumulating only a single partial sum in each PE's register file,
we accumulate $K$ partial sums, one for each LSTM cell being computed concurrently.
With this dataflow, we eliminate the need to fetch $W_{x}$ more than once from DRAM.

\paratitle{PE Array}
\label{sec:arch:mem-x:pe}
\label{sec:arch:mem1-x:pe}
Because Cluster~3 layers have a low MAC intensity,
and mainly perform MVM,
we design a much smaller PE array for \accelB than that in the
baseline accelerator.
We again analyze the inference latency across a range of
PE array sizes, and find that an 8x8 array
(\SI{128}{\giga\flop\per\second} peak throughput)
balances latency, utilization, and energy.

\paratitle{Memory System}
\label{sec:arch:mem-x:mem}
\label{sec:arch:mem1-x:mem}
Since \accelB is data-centric and its sequential
memory accesses exploit high memory bandwidth,
we place it in the logic layer of a
3D-stacked memory, where the bandwidth is much higher
than the external bandwidth to the CPU.
Given that parameters and activations from
 these layers exhibit different characteristics, we 
customize separate on-chip buffers for each data type. 
For parameters, we use only one level of memory hierarchy
(\SI{512}{\byte} private buffer per PE) and stream parameters directly
from DRAM, as (1)~their minimal reuse and large footprint does
not warrant a shared on-chip buffer, (2)~our proposed dataflow 
\sg{now generates sequential accesses that can efficiently utilize
the full memory bandwidth.}
For activations, thanks to the small
activation footprint of layers in Cluster~3, we use a \SI{128}{\kilo\byte} buffer.

\subsection{\accelC Accelerator Design}
\label{sec:arch:mem2-x}

\paratitleattop{}
\paratitle{Dataflow}
\label{sec:arch:mem2-x:dataflow}
We leverage three characteristics of Cluster~4/5 layers
to shape the design of \accelC:
(1)~activations exhibit very low reuse;
(2)~parameters exhibit lower reuse than Cluster~1/2 layers,
but exhibit much higher reuse than activations; and
(3)~the footprint of activations is low for all layers,
while the parameter footprint is high for many (but not all) layers.
Based on these, we need to design a dataflow that exposes
reuse opportunities for parameters, which in turn allows us
to shrink the on-chip parameter buffer.

There are several variations of dataflows that can enable reuse opportunities for parameters. 
For example, we can replicate the same parameter across different PEs,
 and temporally reuse parameter across different iterations.
For each parameter, this results in a reuse factor (i.e., the number of cycles a parameter is reused)
of $WxH/N$, where $N$ is the number of PEs and $WxH$ is the dimension of the output activations. Once we
 traverse all $WxH$ output activations, we need to broadcast
 a new parameter element across the PEs and repeat the process. 
If the output activation dimension 
is small (which is the case for Cluster~4/5 layers), 
the reuse factor for each parameter becomes too small,
resulting in frequent parameter broadcasting.
This is
 particularly problematic for Cluster~4 layers, which often have a large number of
 filters (i.e., a large parameter footprint).
The on-chip traffic generated by frequent parameter broadcasting
prevents us from efficiently overlapping the retrieval of parameters from DRAM
with PE computation, leading to PE underutilization.

To avoid this in \accelC, 
we instead spatially distribute parameters across PEs.
All PEs collaboratively calculate one output activation, while
input activations are spatially distributed among PEs.
The parameters stay in the PE register file
across different cycles (temporal reuse) until we traverse the 
all of the output activations.
This increases the reuse factor to $WxH$ and reduces
on-chip traffic (by eliminating the parameter broadcasts).
With this larger reuse factor, our dataflow can now effectively
hide the DRAM access latency with PE computation time.

\paratitle{PE Array}
\label{sec:arch:mem2-x:pe}
While Cluster~4/5 layers have a low MAC intensity, they perform
more MAC operations on average than Cluster~3 layers. 
Our analysis of PE array sizes shows that 
equipping \accelC with an array smaller than 16x16 increases the latency,
so we select 16x16 for the array size.
This allows \accelC to achieve a peak throughput of
\SI{512}{\giga\flop\per\second}. 

\paratitle{Memory System}
\label{sec:arch:mem2-x:memory}
Similar to \accelB, we 
decide to place \accelC inside the logic layer of 3D-stacked memory.
This provides high bandwidth for the large parameter footprints of Cluster~4 layers.
We use separate shared buffers for each data type. 
Given the small activation footprints, we use
a small \SI{128}{\kilo\byte} buffer for them (a 16x reduction compared to the baseline
 accelerator). Our proposed dataflow for \accelC enables 
\sg{Thanks to the temporal parameter reuse across
 iterations that our dataflow enables,
we can reduce the 
 parameter buffer size by 32x (down to \SI{128}{\kilo\byte}) over the Edge TPU}.


\section{Experimental Methodology}
\label{sec:methodology}

\paratitleattop{}
\paratitle{Models \& Simulation}
\sg{The Google edge NN models that we analyze} are used in several
 Google mobile applications/products, such as image classification, object detection,
 semantic segmentation, automatic speech recognition, and image captioning. The models are
 specifically developed for edge devices using TensorFlow Lite~\cite{tensorflow-lite}, and
 are fully 8-bit quantized 
using \sg{quantization-aware training}~\cite{quant-aware-training}.
 The models are then compiled using the Edge TPU compiler~\cite{edge-tpu-compiler}.
 
\paratitle{Energy Analysis}
We build our energy model based on prior works~\cite{google-pim, tetris, conda}, which sums up the total energy consumed by
the accelerator, DRAM, off-chip and on-chip interconnects, and all on-chip buffers. 
We use CACTI-P 6.5~\cite{CACTI} with a \SI{22}{\nano\meter} process technology
 to estimate on-chip buffer energy. We assume that each 8-bit MAC unit consumes \SI{0.2}{\pico\joule\per\bit}. We model
 the DRAM energy as the energy consumed per bit for LPDDR4, based on models from prior works~\cite{tetris, google-pim}.

\paratitle{Performance Analysis}
We use an in-house simulator \sg{to model all major components of} the Edge TPU,
including the PE array, memory
 system, on-chip network, and dataflow. We heavily modify the simulator to implement our three proposed accelerators
and the software runtime of \titleShort. 
We develop an analytical cost model to determine the performance of each of our proposed dataflows, and integrate
the dataflow performance numbers into our simulator's performance model. We use CACTI-P 6.5~\cite{CACTI}
\sg{to determine the on-chip buffer latencies for each proposed accelerator. 
Similar to prior works~\cite{google-pim, tetris, mondrian, conda}, 
the accelerators in the logic layer of 3D-stacked memory have access to
the \SI{256}{\giga\byte\per\second} internal bandwidth of High-Bandwidth Memory (HBM)~\cite{hbm},
which is 8x the bandwidth to accelerators that sit
outside of memory.}


\section{Evaluation}
\label{sec:eval}

We evaluate energy and performance for \am{four} configurations:
(1)~\emph{Baseline}, the Edge TPU;
(2)~\emph{Base+HB}, a hypothetical version of Baseline with 8x bandwidth (\SI{256}{\giga\byte\per\second}); 
\am{(3)~\emph{EyerissV2}, a state-of-the-art edge accelerator~\cite{eyerissv2} that uses reconfigurable interconnects to address CNN model heterogeneity; and}
(4)~our implementation of \emph{\titleShort} with all three proposed accelerators (\accelA, \accelB, \accelC).
To improve figure clarity, we show individual model results
for only a few representative models of each model type.
Our average results are reported across \emph{all} 24 Google edge models.

\subsection{Inference Energy Analysis}
\label{sec:eval:energy}

Figure~\ref{fig:eval:energy-breakdown} (left) shows the total inference energy across
different NN models by our \am{four} configurations.
We make three observations from the figure.

\begin{figure}[h]
    \vspace{-5pt}
    \centering
        \centering
        \includegraphics[width=\linewidth]{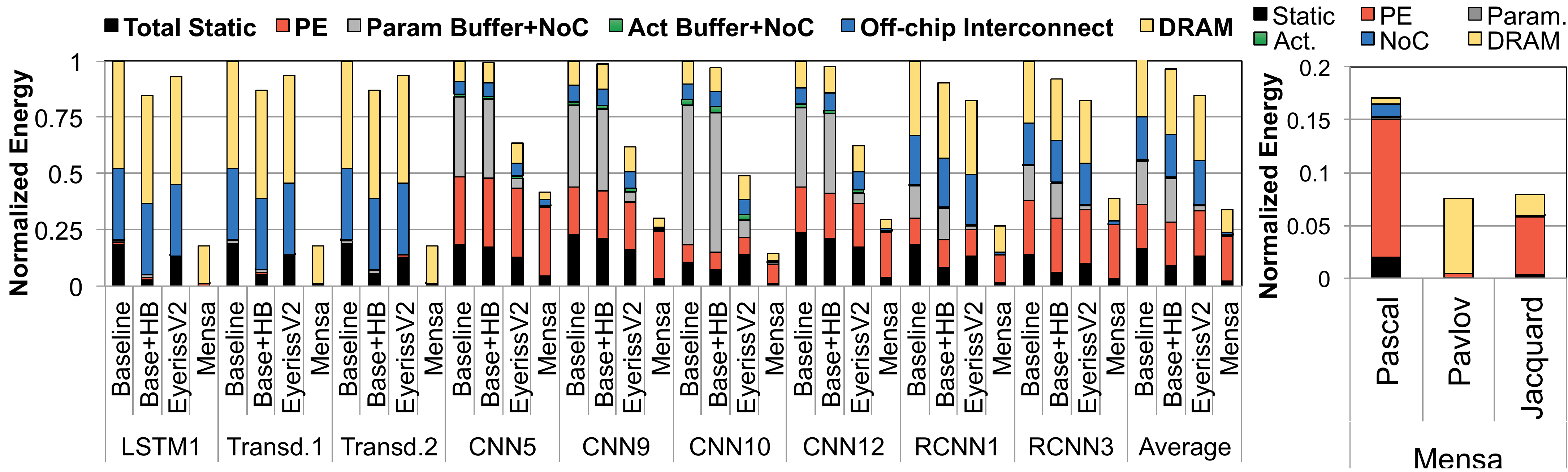}%
\vspace{-22pt}
    \caption{Inference energy across different models (left) and energy breakdown across our three proposed accelerators (right), normalized to Baseline.}
    \label{fig:eval:energy-breakdown}
    \vspace{-5pt}
\end{figure}

First, providing high bandwidth to Baseline results in only a small
reduction (7.5\% on average) of inference energy.
This is because Base+HB still incurs the high energy costs of
(1)~on-chip buffers that are overprovisioned for many layers, and
(2)~off-chip traffic to DRAM.
Base+HB benefits LSTMs and Transducers the most (14.2\% energy reduction),
as the higher bandwidth significantly reduces the inference latency of these models,
which in turn lowers static energy.

\sg{Second, EyerissV2
 suffers from significant energy inefficiency for LSTMs and Transducers. While EyerissV2 lowers static energy
  compared to Baseline, due to its use of a much smaller PE array (384 vs.\ 4096) and on-chip buffers (\SI{192}{\kilo\byte} vs.\ \SI{4}{\mega\byte}), it still
 incurs the high energy costs of off-chip parameter traffic to DRAM. Averaged across all LSTM and Transducer models,
 EyerissV2 reduces energy by only 6.4\% over Baseline. 
For CNN models, EyerissV2 reduces inference energy by 36.2\% over Baseline, as its
smaller on-chip buffer significantly reduces dynamic energy consumption.}

Third, \titleShort significantly reduces inference energy across all models.
The reduction primarily comes from three sources.
(1)~\titleShort lowers the energy spent on on-chip and off-chip parameter traffic
by 15.3x, by scheduling layers on the accelerators with the most appropriate
dataflow for each layer.
LSTMs and Transducers benefit the most, as their inference energy \am{in both Base+HB and EyerissV2}
is dominated by off-chip parameter traffic,
which \accelB and \accelC drastically cut due to being placed inside memory.
(2)~\titleShort reduces the dynamic energy of the on-chip buffer and
network (NoC) by 49.8x \am{and 6.2x over Base+HB and EyerissV2}, by avoiding overprovisioning
and catering to specialized dataflows.
This is most beneficial for CNN and RCNN models.
(3)~\titleShort reduces static energy \am{by 3.6x and 5.6x over Base+HB and EyerissV2}, thanks to
using significantly smaller PE arrays that avoid underutilization,
significantly smaller on-chip buffers, and dataflows that reduce inference latency.

\am{EyerissV2 falls significantly short (50.6\%) of Mensa's energy efficiency for three reasons. First, while
 EyerissV2's flexible NoC can provide a high data rate to the PE array, its fixed dataflow cannot efficiently expose
 reuse opportunities across different layers (e.g., Cluster 4 and 5 layers that have very large parameter
footprints and low data reuse). 
\sg{Second, EyerissV2 has much higher static energy consumption, 
as its inference latency is significantly larger for many compute-intensive CNN layers
(as its PE array is much smaller than the array in \accelA).}
Third, some CNN layers have a large parameter footprint and very low data reuse, which
generates a large amount of off-chip parameter traffic \sg{in EyerissV2}.}
Overall, \titleShort reduces total inference energy by \am{66.0\%/50.6\%, and improves energy efficiency
(\si{\tera\flop\per\joule}) by 3.0x/2.4x, compared to Baseline/EyerissV2}.

Figure~\ref{fig:eval:energy-breakdown} (right) shows the breakdown of
energy usage across our three \titleShort accelerators.
\accelA consumes the most energy of the three, with its
consumption dominated by the PE array (since the layers that run on
\accelA perform a large number of MAC operations).
\accelB's energy usage is dominated by DRAM accesses, as its layers
have large footprints and no data reuse.
For \accelC, the majority of energy is used by a combination of DRAM accesses
and the PE array, but the usage is lower than \accelB DRAM accesses or
the \accelA PE array due to the inherent layer properties (smaller footprints,
lower MAC intensity).

\subsection{Performance Analysis}
\label{sec:eval:performance}

Figure~\ref{fig:eval:utilization} shows the
utilization (bars, left axis) and throughput (lines, right axis)
for our \am{four} configurations.
\titleShort's utilization is calculated by computing
the average utilization across its three accelerators (\accelA, \accelB and \accelC). 
We make three observations from the figure.

\begin{figure}[h]
    \centering
        \centering
        \includegraphics[width=\linewidth]{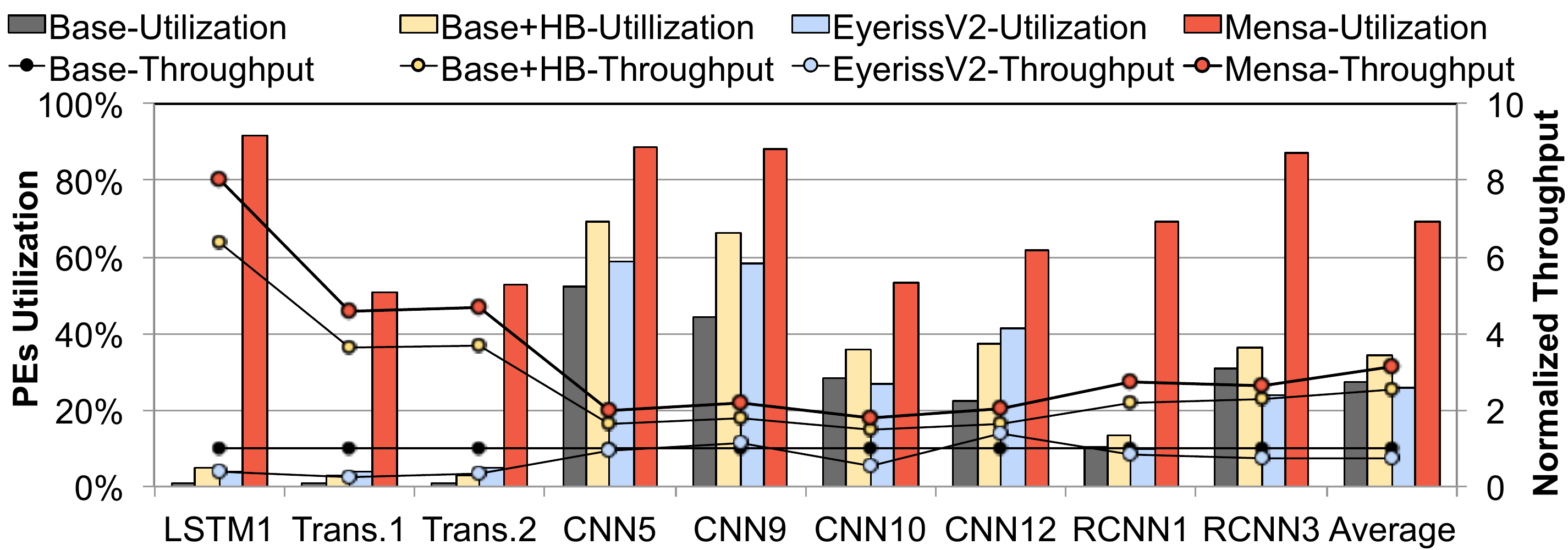}%
\vspace{-22pt}
    \caption{Accelerator utilization and throughput across different models, normalized to Baseline.}
    \label{fig:eval:utilization}
    \vspace{-8pt}
\end{figure}

First, Baseline suffers from low PE array utilization (on average 27.3\%).
The higher bandwidth in Base+HB pushes average utilization up to 34.0\%,
and improves throughput by 2.5x.
The largest improvements are for LSTMs and Transducers (4.5x on average
vs.\ 1.3x for CNNs), thanks to their low
\si{\flop\per\byte} ratio and large footprints.
In contrast, some CNN models (e.g., CNN10) see only modest improvements (11.7\%)
with Base+HB, as their layers have high reuse and small footprints.
Overall, Base+HB still has very low utilization, as many
layers (those from Clusters 3, 4, and 5) do not need the large number of
PEs in the accelerator. 

\sg{Second, EyerissV2 \emph{reduces} performance significantly over Baseline
for several models.  EyerissV2's flexible interconnect and much 
 smaller PE array do allow it to achieve slightly higher PE utilization than Baseline for layers that have very low data
 reuse. 
 However, this higher utilization is offset by significantly higher inference latencies.
For compute-intensive layers in Clusters 1 and 2, the smaller PE array size hurts layer throughput.
For data-centric layers in Clusters 4 and 5, EyerissV2 cannot customize its dataflow
to expose reuse opportunities, and thus is hurt
by the high off-chip traffic.
Overall, we find that EyerissV2's overall throughput is actually \emph{lower}
than Baseline for most of our models.}

\am{Third, \titleShort provides a significant increase in both average utilization
(2.5x/2.0x/2.6x) and throughput
(3.1x/1.3x/4.3x) over Baseline/Base+HB/EyerissV2.}
The large utilization improvements are a result of
(1)~properly-provisioned PE arrays for each layer,
(2)~customized dataflows that exploit reuse and opportunities for parallelization, and
(3)~the movement of large-footprint layer computation into memory
(eliminating off-chip traffic for their DRAM requests).
We note that \titleShort's throughput improvements over Base+HB
are smaller than its utilization improvements,
because Base+HB is reasonably effective at reducing the inference latency
of layers with poor reuse and large footprints
(albeit with poor energy efficiency and
underutilization).
\titleShort benefits all NN model types, but the largest improvements are for
LSTMs and Transducers, with average utilization/throughput improvements of
82.0x/5.7x over Baseline.
The improvement is lower for CNNs and RCNNs (2.23x/1.8x over Baseline),
because they make more use of Baseline's large PE arrays, and
have smaller footprints that lessen the impact of off-chip DRAM accesses.
For a few CNNs (CNN10--CNN13),
their utilization is somewhat lower than desired (44.7\%) due to their use
of a large number of depthwise layers (part of Cluster~5) that have 
significantly lower data reuse than other Cluster~4/5 layers.
While these layers run less optimally with \accelC's dataflow due to the
different reuse behavior, \titleShort still improves their utilization
by 65.2\% over Baseline because \accelC's specialization
still helps depthwise layers.

Figure~\ref{fig:eval:latency} shows the overall inference latency,
and where the inference latency in \titleShort is spent (across \accelA, \accelB, and \accelC).
We find that \titleShort outperforms Baseline and Base+HB on average by 1.96x and 1.17x.
LSTMs and Transducers see a significant latency reduction with \titleShort
(5.4x/1.26x vs.\ Baseline/Base+HB) because most of their layers
run on \accelB and benefit from an optimized dataflow and processing-in-memory
(which provides not only higher bandwidth, but also lower latency for DRAM accesses).
CNNs and RCNNs benefit from the heterogeneity of our accelerators,
making use of all three of them to reduce latency by 1.64x/1.16x over Baseline/Base+HB.

\begin{figure}[h]
    \vspace{-5pt}
        \centering
        \includegraphics[width=\linewidth]{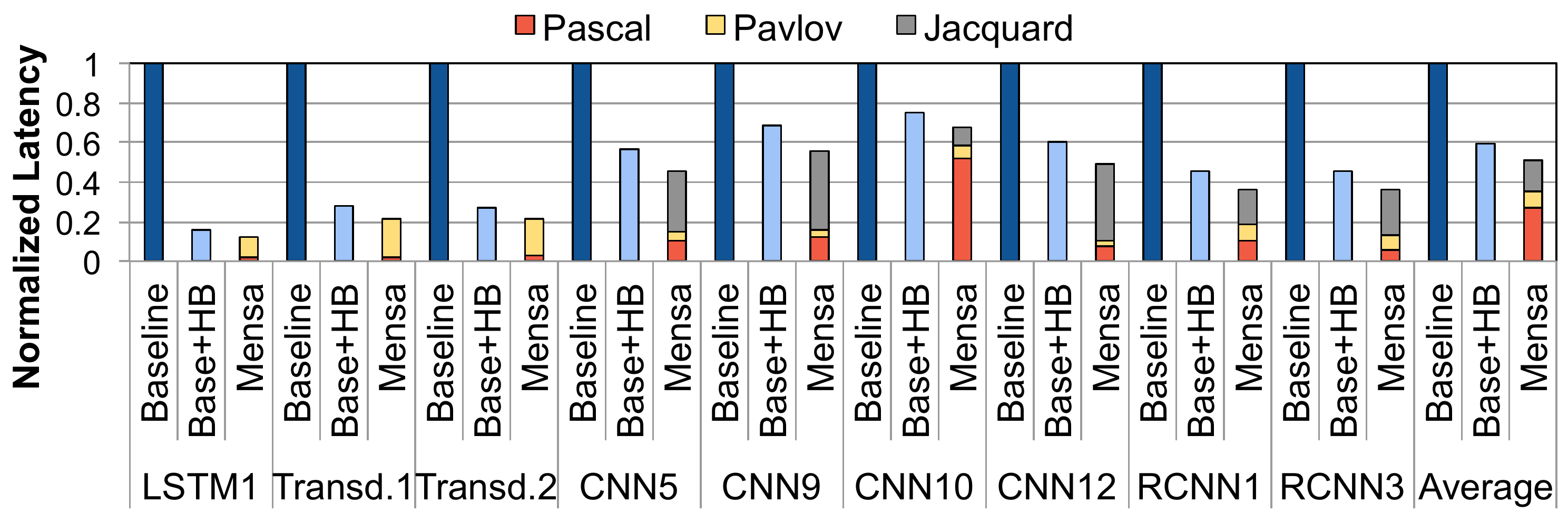}%
\vspace{-22pt}
    \caption{Inference latency, normalized to Baseline.}
    \label{fig:eval:latency}
    \vspace{-10pt}
\end{figure}


\section{Related Work}
\label{sec:related}

\sg{To our knowledge, this is the first work to
(1)~examine the bottlenecks of a state-of-the-art commercial
Google Edge TPU when executing state-of-the-art Google edge NN
models;
(2)~quantify the significant layer variation that exists in state-of-the-art
edge NN models;
(3)~identify that layers can be clustered together based on a number of
shared characteristics;
(4)~propose a new framework for heterogeneous ML inference acceleration (\titleShort),
with both on-chip and near-data accelerators; and
(5)~provide an example heterogeneous accelerator design for
Google edge NN models.}

Many prior works~\cite{e-pur, eyeriss, tetris, eyerissv2, shidiannao, serving-rnn, 
masr, scaledeep, rana, wax, cnn-resource, recnmp, cnvlutin, fused-layer, 
diannao, flexflow, scnn, simba} look at a specific model type
(predominantly CNNs).
None of these works
perform an analysis across different classes of edge models (e.g., Transducers, LSTMs, RCNNs).
In fact, many of them (e.g., \cite{eyeriss, 
tetris, shidiannao, simba, scnn}) analyze traditional models (e.g., AlexNet, VGG),
and their proposals are not tailored toward state-of-the-art edge models (which we show are different)
or resource-limited edge devices. These proposals cater accelerators
toward a particular model type (e.g., CNNs~\cite{eyeriss, tetris, rana, cnn-resource, scnn},
 LSTMs~\cite{e-pur,serving-rnn}), and they are not optimized to serve multiple model types. 
As a result, they all suffer from the issues we discuss in Section~\ref{sec:motiv}.

Some CNN-focused works observe diversity
across CNN layers~\cite{meastro, scaledeep, eyerissv2, maeri, neurosurgeon, simba}.
Maeri~\cite{maeri} and Eyeriss~v2~\cite{eyerissv2}
 use reconfigurable logic 
 to provide flexible dataflows and networks for different layers.
However, those solutions 
\sg{(1)~cannot reconfigure a number of essential
design parameters,
(2)~require frequent online reconfiguration to cater to edge layer
variation, and
(3)~make it difficult to co-design the dataflow with key components such as the memory system.}
\sg{We quantitatively compare Eyeriss~v2 to \titleShort in Section~\ref{sec:eval}.}
ScaleDeep~\cite{scaledeep}
 proposes customized processing tiles to address diversity in CNN layers. While the idea shares
some similarities with \titleShort (exploiting heterogeneity in hardware),
ScaleDeep (1)~targets traditional cloud-based training instead of edge inference,
resulting in significantly different support for diversity; and
(2)~does not address the larger diversity that exists between CNN layers
and layers in other model types (e.g., LSTM layers).
Neurosurgeon~\cite{neurosurgeon} looks at both vision and speech models. However, 
their analysis is done on old/traditional vision/speech models,
and 
their solution relies on offloading some layers to the cloud, which undermines the goal of running inference locally.


\section{Conclusion}
\label{sec:conclusion}

\sg{We conduct the first bottleneck analysis of the
Google Edge TPU, a state-of-the-art ML inference accelerator,
as it executes 24 state-of-the-art Google edge NN models.
Our analysis reveals that 
the Edge TPU's monolithic design leads to
significant underutilization and poor energy efficiency for our
edge NN models, which exhibit significant layer variation.
We propose a new framework called \titleShort,
consisting of multiple small heterogeneous accelerators,
each specialized to specific layer characteristics.
Using our discovery that layers group into
a small number of clusters, we create a \titleShort design
for the Google edge NN models consisting of three
accelerators.
Compared to the Edge TPU and to a state-of-the-art
reconfigurable ML accelerator,
our design improves energy efficiency by 3.0x/2.4x and
throughput by 3.1x/4.3x for our edge NN models.
We hope that \titleShort can enable the design and
adoption of future heterogeneous accelerators that support
yet-to-be-developed NN model types.}

{
\interlinepenalty=10000
\bibliographystyle{IEEEtranS}
\bibliography{references}
}

\end{document}